\documentclass[a4paper,fleqn,usenatbib]{mnras}

\usepackage{newtxtext,newtxmath}

\usepackage[T1]{fontenc}
\usepackage{ae,aecompl}

\usepackage{graphicx}	
\usepackage{amsmath}	
\usepackage{amssymb}	

\title[Quasars dominated by nuclear dust emission]{The VMC Survey -- XLIX. Discovery of a population of quasars dominated by nuclear dust emission behind the Magellanic Clouds}

\author[C. M. Pennock et al.]{Clara M. Pennock,$^{1}$\thanks{E-mail: c.m.pennock@keele.ac.uk}
Jacco Th. van Loon,$^{1}$ 
Joy O. Anih,$^{1}$
Chandreyee Maitra,$^{2}$ 
\newauthor Frank Haberl,$^{2}$ 
Anne E. Sansom,$^{3}$
Valentin D. Ivanov,$^{4}$
Michael J. Cowley,$^{5,6}$
\newauthor Jos\'e Afonso,$^{7,8}$ 
Sonia Ant\'{o}n,$^{9}$
Maria-Rosa L. Cioni,$^{10}$
Jessica E. M. Craig,$^{1}$
\newauthor Miroslav D. Filipovi\'c,$^{11}$
Andrew M. Hopkins,$^{12}$
Ambra Nanni,$^{13}$
Isabella Prandoni,$^{14}$
\newauthor Eleni Vardoulaki$^{15}$
\\
$^{1}$Lennard-Jones Laboratories, Keele University, ST5 5BG, UK\\
$^{2}$Max Planck Institute for Extraterrestrial Physics, Gie{\ss}enbachstra{\ss}e, 85748 Garching bei M{\"u}nchen, Germany\\
$^{3}$Jeremiah Horrocks Institute, University of Central Lancashire, Preston PR1 2HE, UK\\
$^{4}$European Southern Observatory, Karl-Schwarzschild-Str.\ 2, D-85748 Garching bei M\"unchen, Germany\\
$^{5}$School of Chemistry and Physics, Queensland University of Technology (QUT), 2 George Street, Brisbane, QLD 4000, Australia\\
$^{6}$Centre for Astrophysics, University of Southern Queensland, West St, Darling Heights, Toowoomba, QLD 4350, Australia\\
$^{7}$Instituto de Astrof\'{\i}sica e Ci\^{e}ncias do Espa\c co, Universidade de Lisboa, OAL, Tapada da Ajuda, PT1349-018 Lisboa, Portugal\\
$^{8}$Departamento de F\'{\i}sica, Faculdade de Ci\^{e}ncias, Universidade de Lisboa, Edif\'{\i}cio C8, Campo Grande, PT1749-016 Lisbon, Portugal\\
$^{9}$CFisUC, Departamento de F\'{\i}sica, Universidade de Coimbra, 3004-516 Coimbra, Portugal\\
$^{10}$Leibniz-Institut f\"ur Astrophysik Potsdam, An der Sternwarte 16, D-14482 Potsdam, Germany\\
$^{11}$Western Sydney University, Locked Bag 1797, Penrith South DC, NSW 2751, Australia\\
$^{12}$Australian Astronomical Optics, Macquarie University, 105 Delhi Rd., North Ryde, NSW 2113, Australia\\
$^{13}$National Centre for Nuclear Research, ul.\ Pasteura 7, 02-093 Warszawa, Poland\\
$^{14}$INAF - Istituto di Radioastronomia, Via P. Gobetti 101, 40129, Italy\\
$^{15}$Th\"{u}ringer Landessternwarte, Sternwarte 5, 07778 Tautenburg, Germany\\
}

\date{Accepted XXX. Received YYY; in original form ZZZ}

\pubyear{2022}

\begin{document}
\label{firstpage}
\pagerange{\pageref{firstpage}--\pageref{lastpage}}
\maketitle

\begin{abstract}
Following the discovery of SAGE0536AGN ($z \sim$ 0.14), with the strongest 10-$\mu$m silicate emission ever observed for an Active Galactic Nucleus (AGN), we discovered SAGE0534AGN ($z \sim$ 1.01), a similar AGN but with less extreme silicate emission. Both were originally mistaken as evolved stars in the Magellanic Clouds. Lack of far-infrared emission, and therefore star-formation, implies we are seeing the central engine of the AGN without contribution from the host galaxy. They could be a key link in galaxy evolution.
We used a dimensionality reduction algorithm, t-SNE (t-distributed Stochastic Neighbourhood Embedding) with multi-wavelength data from Gaia EDR3, VISTA survey of the Magellanic Clouds, AllWISE and the Australian SKA Pathfinder to find these two unusual AGN are grouped with 16 other objects separated from the rest, suggesting a rare class. Our spectroscopy at SAAO/SALT and literature data confirm at least 14 of these objects are extragalactic ($0.13 < z < 1.23$), all hosting AGN. Using spectral energy distribution fitter C\textsc{igale} we find that the majority of dust emission ($> 70 \%$) in these sources is due to the AGN. Host galaxies appear to be either in or transitioning into the green valley. There is a trend of a thinning torus, increasing X-ray luminosity and decreasing Eddington ratio as the AGN transition through the green valley, implying that as the accretion supply depletes, the torus depletes and the column density reduces. Also, the near-infrared variability amplitude of these sources correlates with attenuation by the torus, implying the torus plays a role in the variability.

\end{abstract}




\section{Introduction}




Active Galactic Nuclei (AGN) preside in the centre of some galaxies, resulting from the accretion of gas by a supermassive black hole. The mass of supermassive black holes is known to correlate with the mass of the galaxy bulge, implying the formation and evolution of bulges and supermassive black holes are intertwined \citep{1998AJ....115.2285M,2000ApJ...539L...9F,2000ApJ...539L..13G,2002ApJ...574..740T}. It is thought that AGN play a significant role in galaxy evolution by creating large outflows that quench \citep{2013ARA&A..51..511K} and/or trigger bursts of star formation \citep{2010ApJ...720..368X,2011MNRAS.414.1082M,2015A&A...573A..85R,2016MNRAS.457..629C}, making them an ideal laboratory for studying the evolution and formation of galaxies. 

AGN emit across the entire electromagnetic spectrum. The diversity of observed AGN can be explained by a small number of physical parameters, such as the mass of the central supermassive black hole (SMBH), the rate of gas accretion onto the black hole, the orientation of the accretion disk with respect to our line-of-sight, the degree of obscuration of the nucleus by dust, and the presence or absence of jets. This is called the unified model of AGN \citep{1993ARA&A..31..473A,1995PASP..107..803U}. This model is however an oversimplification of observed variety of AGN evolving through cosmic time \citep[see][]{2016MNRAS.457..629C,2018MNRAS.473.3710C}. Finding the more unusual of these diverse objects could be the key to unlocking the evolution of AGN, such as an AGN without interstellar gas to feed it or AGN hosted by bulgeless galaxies \citep[e.g.][]{2017MNRAS.470.1559S}, implying no history of major mergers.

Emission from hot dust is associated with the torus of gas and dust surrounding the central engine of the AGN and most often observed in the mid-infrared (mid-IR) \citep{1982Natur.299..605A,1984ApJ...278..499A,1988ApJ...325...74S}. The distribution (smooth, clumpy or polar) and kinematics (static, inflowing or outflowing) of this hot dust are however still uncertain. For instance, at parsec scales in the polar regions there exist grains, thought to be irradiated by the AGN almost directly \citep{Raban2009,2012ApJ...755..149H,2013ApJ...771...87H,2014A&A...563A..82T,2016ApJ...822..109A,2016A&A...591A..47L,2018ApJ...862...17L,2019ApJ...884..171H}, which may be associated with an AGN-driven outflow \citep{2014MNRAS.445.3878S}. The properties of these grains observed in the torus and polar regions appear to be different from those observed in the interstellar medium (ISM), with a dearth of smaller grains such as small graphite grains and/or polycyclic aromatic hydrocarbon (PAH) nanoparticles, indicated by the absence of a 2175 \AA\ bump \citep{2004MNRAS.348L..54C,2004ApJ...616..147G,2007arXiv0711.1013G}, whilst retaining larger grains such as silicate.

Silicate features in emission are expected for AGN seen face-on (type 1 AGN), where dust in the surface of the inner torus will be heated by radiation from the central engine to sufficient temperatures to allow for direct detection of the 10 $\mu$m and 18 $\mu$m silicate bands emitted from this hot dust. The Spitzer space telescope has been used to detect this emission in type 1 AGN \citep{2005ApJ...625L..75H,2005AN....326R.556S,2005ApJ...629L..21S,2005ApJ...633..706W,2006ApJ...653..127S,2015ApJ...803..110H} as well as in type 2 AGN \citep{2007ApJ...655L..77H}, where it would be expected to be detected in absorption. Silicate emission detected in type 2 AGN breaks the relation between orientation and AGN characteristics; this is explained by clumpiness in the torus seen in the radiative transfer models of \cite{2008ApJ...685..147N} and \cite{2009ApJ...707.1550N}.

The originator of the strongest 10 $\mu$m silicate emission of any known AGN, is the hot dust near the SMBH of SAGE1C J053634.78$-$722658.5 (hereafter referred to as ‘SAGE0536AGN’) that was discovered serendipitously behind the LMC by \cite{2011A&A...531A.137H} in the Spitzer Space Telescope Survey of the Agents of Galaxy Evolution Spectroscopic follow-up of IR sources seen towards the LMC \citep[SAGE-Spec: ][]{2010PASP..122..683K,2011MNRAS.411.1597W}. It lies behind the Large Magellanic Cloud (LMC) and was found to be a type 1 AGN with a negligible amount of far-IR emission meaning a lack of star formation, confirmed by spectra obtained with the Southern African Large Telescope (SALT) \citep{2015MNRAS.453.2341V}. Finding more of these could provide valuable insight into this stage of galaxy/AGN evolution.

Our new spectroscopic observations using the South African Astronomical Observatory (SAAO) 1.9m telescope, reveals SSTISAGE1C J053444.17$-$673750.1 is one such source that shows similarities to SAGE0536AGN.  This source has also been referred to as 4XMM J053444.1$-$673751, 2MASS J05344418$-$6737501, SHP LMC 256 or [KWV2015] J053444.17$-$673750.1 (identifier for post-AGB star candidate), in this paper it shall be referred to as SAGE0534AGN. Both of these sources have been confused as evolved stars, have silicate emission and a lack of star formation. Can more be found? Are they an unusual type, or a short and therefore rarely seen stage of galaxy/evolution?

As these sources mimic evolved stars in the Magellanic Clouds, we therefore needed to adopt a more systematic approach in finding more of them. Unsupervised machine learning has been used to great effect to cluster objects together and reveal patterns in large datasets \citep[e.g.,][]{2016ApJS..225...31L,2018A&A...619A.125A,2018MNRAS.476.2117R,2020ApJ...905...97Z}. This can be used to find objects with similar properties to those already discovered, such as SAGE0536AGN and SAGE0534AGN.

AGN are most readily identified within combinations of multi-wavelength photometric survey data. The Magellanic Clouds span $\sim$ 100 sq.\ degrees on the sky that have been studied, in parts or as a whole, in the UV \citep[e.g.][]{2014AAS...22335511T}, optical \citep[e.g. Gaia, SMASH;][]{Gaia2021,2017AJ....154..199N}, IR \citep[e.g. SAGE, AllWISE;][]{2004Lacy,2014yCat.2328....0C}, radio \citep[e.g. MOST, ATCA;][]{Mauch2003,Murphy2010} and X-ray \citep[XMM-Newton;][]{2013XMM}, which makes them an ideal location to search for AGN behind them. The combination of all these data has great potential for discovery of the more unusual and extreme cases of AGN, such as SAGE0536AGN. The new and deeper surveys towards the Magellanic Clouds, such as the near-IR VISTA Magellanic Clouds \citep[VMC;][]{2011A&A...527A.116C} and radio Evolutionary Map of the Universe all-sky \citep[EMU;][]{2019MNRAS.490.1202J,2021MNRAS.506.3540P} surveys greatly enhance such attempts.

The paper is laid out as follows: Section \ref{datasection} describes the data used and machine learning tool used to create the sample. In Section \ref{resultssection} we describe the light curves (Section \ref{lightcurvesection}) and spectra, calculation of black hole masses (Section \ref{spectrasection}), spectral energy distribution (SED) fitting with C\textsc{igale} (Section \ref{cigalesection}), X-ray observations (Section \ref{xraysection}) and modelling of those sources where we can see their host galaxies with G\textsc{alfit} (Section \ref{galfitsection}). In Section \ref{discussionsection} we discuss the selection techniques of AGN (Section \ref{selectionsection}) and where this sample and sources mistaken for AGN fall within them. This is followed by a discussion of the sample galaxies' identity as either star-forming, quiescent or green valley galaxy and how their properties change as they transition from star-forming to green valley (Section \ref{greenvalleysection}). The radio properties and how they link to the evolutionary stage of the sample are then discussed (Section \ref{radiosection}) followed by a discussion of the AGN dust and its effect on observed properties such as variability and the 10 $\mu$m silicate emission (Section \ref{dustsection}). 

\section{The Data}\label{datasection}

\subsection{Photometry}

\subsubsection{VISTA Magellanic Clouds Survey}

The VISTA Survey of the Magellanic Clouds \citep[VMC;][]{2011A&A...527A.116C} is a near-IR deep, multi-epoch and wide-field study of the Magellanic Clouds. It has a spatial resolution of $< 1^{\prime\prime}$ in the $YJK$\textsubscript{s} filters, reaching a sensitivity of about 21 mag (Vega). Its depth and coverage can be compared to the VISTA Deep Extragalactic Observations \citep[VIDEO; ][]{2013MNRAS.428.1281J} survey, which was specifically designed to study galaxy and cluster/structure evolution. The VMC data provide an opportunity to double the effort of the VIDEO survey and cover more volume and cosmic variance, and has already proven successful in discovering more AGN \citep[e.g.][]{2016A&A...588A..93I}. This however comes with the caveat of increased stellar confusion with the presence of the LMC and SMC.

\subsubsection{Radio EMU-ASKAP survey}

The Australian Square Kilometre Array Pathfinder (ASKAP) observed the LMC at 888 MHz \citep[54,612 sources;][]{2021MNRAS.506.3540P} with a bandwidth of 288 MHz and beam size of $13\rlap{.}^{\prime\prime}9\times12\rlap{.}^{\prime\prime}1$, and the SMC at 960 MHz and 1320 MHz \citep[7,736 sources;][]{2019MNRAS.490.1202J} with a bandwidth of 192 MHz and beam sizes of $30^{\prime\prime}\times30^{\prime\prime}$ and $16\rlap{.}^{\prime\prime}3\times15\rlap{.}^{\prime\prime}1$, respectively. The majority of these sources were found to be extragalactic.


\subsubsection{Other survey data}

Other data used in this work include optical Gaia EDR3 survey \citep{Gaia2021} photometry and astrometry; the optical Survey of the Magellanic Stellar History \citep[SMASH;][]{2017AJ....154..199N} photometry; mid-IR AllWISE \citep{2014yCat.2328....0C} and Spitzer SAGE \citep{2004Lacy} photometry and XMM-Newton \citep{2013XMM} X-ray photometry.

\subsection{Spectroscopy}
\subsubsection{SAAO 1.9m spectra}

Eight new optical spectra of SAGE0534AGN and Source 1, 2, 3, 4, 6, 11 and 14 (see Table \ref{tab:tsnesample} for sample list) were obtained at the South African Astronomical Observatory (SAAO) 1.9m telescope with SpUpNIC \citep[Spectrograph Upgrade: Newly Improved Cassegrain;][]{2019JATIS...5b4007C}. Grating 7 (grating angle of $16^\circ$) and the order blocking ‘BG38’ filter were used, delivering a resolving power $R = \frac{\lambda}{\Delta\lambda}\sim$ 500 over a wavelength range of 3800 \AA\ -- 9000 \AA. The CuAr lamp was used for wavelength calibration. Three 600s exposures were obtained for each source. The standard stars \citep[EG 21, Feige 110 or LTT 1020;][]{1994PASP..106..566H} were observed on the same night under the same conditions for 30s. The data were processed using the standard IRAF\footnote{IRAF is distributed by the National Optical Astronomy Observatory, which is operated by the Association of Universities for Research in Astronomy, Inc., under cooperative agreement with the National Science Foundation.} tools \citep{1986SPIE..627..733T,1993ASPC...52..173T}.

\subsubsection{SALT spectra}

Supplementary optical spectroscopic observations were made of three sources (SAGE0534AGN and Source 13 and 16, see Table \ref{tab:tsnesample} for sample list) using SALT \citep{2006SPIE.6267E..0ZB} under programmes 2021-1-SCI-018 (PI: Jacco van Loon), 2021-1-SCI-029 (PI: Jacco van Loon), 2021-1-SCI-032 (PI: Jacco van Loon) and 2021-2-SCI-017 (PI: Joy Anih). We used the Robert Stobie Spectrograph \citep[RSS;][]{2003SPIE.4841.1463B, 2003SPIE.4841.1634K}, a combination of three CCD detectors with total 3172 $\times$ 2052 pixels and spatial resolution of $0\rlap{.}^{\prime\prime}1267$ per pixel. We used the long-slit with width $1\rlap{.}^{\prime\prime}5$, grating PG0300 and an Argon arc lamp. These data were also processed using the standard IRAF tools \citep{1986SPIE..627..733T,1993ASPC...52..173T}.

Prior to this study SAGE0536AGN had been observed with SALT by \cite{2015MNRAS.453.2341V}. Further observations of SAGE0536AGN were obtained with SALT RSS in 2017 (programme 2017-1-SCI-001) but were unfortunately affected by focus issues. These spectra covered $\sim$534 to 623 nm, with PG2300 grating, including Hb, Mgb and Fe5335 spectral features. Two of the five exposures (observed on 20/10/2017) were of sufficiently good quality and high spectral resolution to attempt kinematic measurements. Using Python PPXF\footnote{\url{https://www-astro.physics.ox.ac.uk/~cappellari/software/\#ppxf}} and INDO-US star spectral templates \citep{2004ApJS..152..251V} the measured velocity dispersion was $s\sim$ 202 $\pm$ 15 km s$^{-1}$, with overall errors from PPXF uncertainty and spectral resolution uncertainty added in quadrature. This measurement was within the central $\sim1^{\prime\prime}$ along the major axis of SAGE0536AGN and is larger than previously found, $s\sim$ 123 $\pm$ 15 km s$^{-1}$, in \cite{2015MNRAS.453.2341V}. This may be because of the focus problems with the 2017 data but could also result from measurement in a better spectral range, less affected by a particular (NaD) spectral feature and along the major axis. IFU data would be needed to more accurately determine the kinematics across SAGE0536AGN.

\subsubsection{Other optical spectra}

Prior to this study, three sources (Source 7, 9 and 10, see Table \ref{tab:tsnesample} for sample list) had been observed as part of the Magellanic Quasars Survey \citep[MQS,][]{2013ApJ...775...92K} and one source (Source 12) as part of a search for variability-selected quasars in the Magellanic Field \citep{2003AJ....125....1G}.

Another (Source 15) had been observed with European Southern Observatory's 3.6m telescope with EFOSC2 as part of a survey to find polarized quasars \citep[see][ Kishimoto et al. \textit{in prep.}]{2008Natur.454..492K}. For all frames, the CCD was read out with 2 $\times$ 2 binning, giving a
spatial sampling of $0\rlap{.}^{\prime\prime}316$ per pixel. The grism Gr\#1 was used at
a dispersion of 13 \AA\ per pixel (after the binning).  The target was
observed with $1\rlap{.}^{\prime\prime}5$ slit width, giving a spectral resolution of
$\sim60$ \AA. The data were reduced in a standard manner.
Averaged bias frame was subtracted, and each frame was flat-fielded.
The wavelengths were calibrated using arc frames, and the spectra
were extracted with $2\rlap{.}^{\prime\prime}8$ window and flux-calibrated.

\subsection{Target sample}

\subsubsection{SAGE0536AGN}

SAGE1C J053634.78$-$722658.5 is a peculiar and rare AGN at $z =$ 0.14 \citep{2015MNRAS.453.2341V}, which was discovered serendipitously, as its colours indicated it was most likely a dusty evolved star. It was discovered by \cite{2011A&A...531A.137H} in the SAGE-Spec survey \citep{2010PASP..122..683K,2011MNRAS.411.1597W} to be an AGN and further characterised by \cite{2015MNRAS.453.2341V}. 

\subsubsection{SAGE0534AGN}

This source was spectroscopically observed as part of SAGE-Spec \citep{2011MNRAS.411.1597W}. The classification was based on a combination of infrared spectral features, continuum and spectral energy distribution shape, bolometric luminosity, cluster membership and variability information. It was described as an unusual object as it demonstrates a very broad 20 $\mu$m emission and a very broad but weak 10 $\mu$m emission. It also has a double peaked SED, which is often taken as indicative of a post-Asymptotic Giant Branch (AGB) object, though it was considered bluer than expected for a post-AGB object. This study also considered X-ray counterparts. It was identified as an X-ray source of unknown physical nature by \cite{2000A&AS..143..391S}, source ID 256, when it was observed by the ROSAT High Resolution Imager \citep[HRI;][]{1995SPIE.2518...96Z}, as well as detected by the X-ray Multi-Mirror Mission \citep[XMM/2XMMi;][]{2009A&A...493..339W} where it showed an SED that peaks around 1 keV. This, combined with the unusual Spitzer IRS spectrum, led to a classification of 'Unknown'. This source was also detected five times serendipitously in the field of view of XMM-Newton observations \citep{2020A&A...641A.136W,2022yCat.9065....0W} and is designated as 4XMM J053444.1$-$673751.

SAGE0534AGN was first spectroscopically observed in the optical as part of a search for optically bright post-AGB stars in the LMC \citep{2011A&A...530A..90V}. On the basis of a low resolution spectrum this object was determined to be a post-AGB star of spectral-type G. A later optical spectroscopic study by \cite{2015MNRAS.454.1468K} revealed this source to be a quasi-stellar object (QSO) instead. As this source was not stellar in nature it was not further explored in that study and the spectrum was not published.

\begin{figure}
	\includegraphics[trim={0.5cm 2.4cm 0cm 1.2cm}, width=\columnwidth]{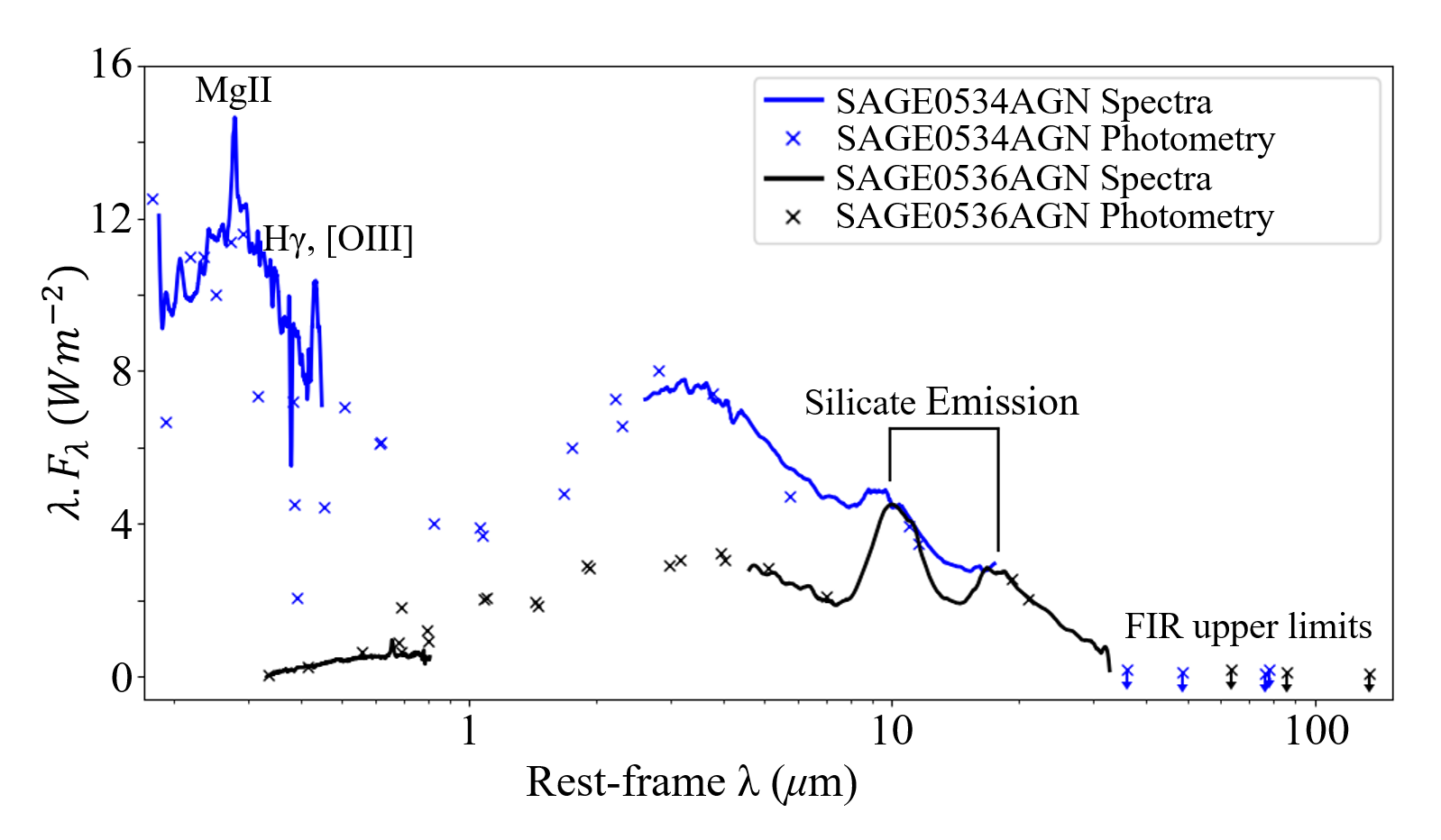}
    \caption{Comparison of SEDs from optical to far-IR of SAGE0536AGN and SAGE0534AGN. Both show the 1 -- 10 $\mu$m bump that is associated with AGN, a noticeable 10 $\mu$m silicate emission, as well as a lack of far-IR emission.}
    \label{fig:SED_comp}
\end{figure}

Comparison of the full SEDs of SAGE0536AGN and SAGE0534AGN is shown in Figure \ref{fig:SED_comp}. From this we can see that they share a lack of far-IR emission, indicating a lack of emission from star formation. At the optical/UV end there is a lack of emission for SAGE0536AGN, whereas SAGE054AGN is bright, implying more dust extinction in SAGE0536AGN. We can also see the silicate emission, which is much stronger for SAGE0536AGN than SAGE0534AGN.

\subsubsection{Machine learning}

In order to find sources similar to SAGE0536AGN and SAGE0534AGN in a large dataset with ill defined properties, we employ machine learning, which has been used to great effect to separate sources into different classes \citep[e.g.,][]{2016ApJS..225...31L,2018A&A...619A.125A,2018MNRAS.476.2117R,2020ApJ...905...97Z}.

T-SNE \citep[t-distributed stochastic neighbour embedding;][]{2008tsne} is an unsupervised machine learning dimensionality reduction algorithm. It can visualise any high–dimensional dataset by projecting each data-point onto a low–dimensional map, which reveals local as well as global structure of the data at many diﬀerent scales. T-SNE has been shown to be adept at separating sources into different classes with no prior information about the source nature \citep[e.g. ][]{2020ApJ...891..136S}.

T-SNE uses hyperparameters (perplexity, early exaggeration, learning rate and number of steps) and is a non-linear technique through non-deterministic or randomised algorithm. It embeds the points from a higher dimension into a lower dimension whilst trying to maintain the neighbourhood of that point, preserving the local structure of the data. More specifically, the t-SNE technique minimizes the divergence between a probability distribution that measures pairwise similarities of the high-dimensional data and a probability distribution that measures pairwise similarities of the low-dimensional points in the embedding. Unlike the linear Principal Component Analysis (PCA) algorithm, t-SNE cannot preserve global structure (variance) but can preserve the local structure, allowing fine structures to be found, which PCA is incapable of.

\begin{figure*} 
\centering
\begin{tabular}{c}
	\includegraphics[trim={1cm 1cm 0cm 0cm},width=\textwidth]{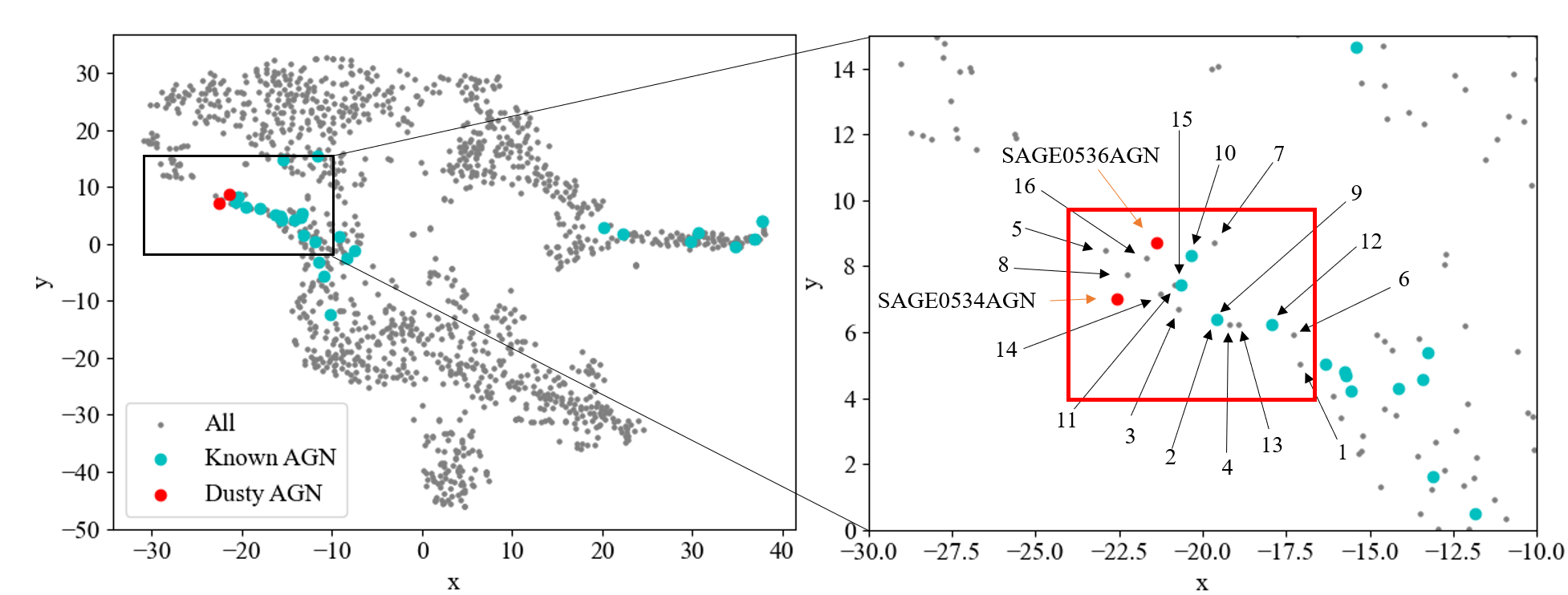}
	\end{tabular}
    \caption{(Left) t-SNE map created from a combination of VMC, Gaia EDR3, AllWISE and ASKAP data. Blue dots represent known AGN that have been spectroscopically confirmed. Red indicates the dusty AGN SAGE0536AGN and SAGE0534AGN. (Right) Zoom in on area containing the dusty AGN. The 16 sources (blue and grey dots) within the red box in this plot represent the sample explored in this paper. The numbers correspond to the source names in Table \ref{tab:tsnesample}. }
    \label{fig:tsnemap}
\end{figure*}

We searched for SAGE0536AGN and SAGE0534AGN analogues to further explore this AGN class. We used the t-SNE algorithm on a clean dataset (no error/missing values) of 1,359 sources that was the combination of VMC, Gaia EDR3 \citep{Gaia2021}, AllWISE \citep{2014yCat.2328....0C} and EMU ASKAP 960 MHz \citep{Joseph2019} and 888 MHz \citep{2021MNRAS.506.3540P} photometry, colours and astrometry in the area of the SMC. Surveys of the SMC have also been performed in the X-ray \citep[e.g.][]{2013XMM}, UV \citep[GALEX;][]{2005GALEX} and mid to far-IR \citep[SAGE/HERITAGE][]{2006SAGE,2013HERITAGE}. These were not used because they lack the same coverage of the Magellanic Clouds as the VMC survey, as well as having missing values for many of observed sources, which would have caused the sample to be explored to be reduced significantly. 
We focus on the SMC because the VMC Point-Spread Function (PSF) photometry and ASKAP radio survey were available for the SMC first.  This technique reduced the high-dimensional dataset down to two dimensions, producing a t-SNE map seen in Figure \ref{fig:tsnemap}. The perplexity parameter of t-SNE was chosen by creating multiple maps and choosing the value of perplexity that created the most obvious clustering.

SAGE0536AGN and SAGE0534AGN are shown to be close to each other in an area containing few sources, implying a rare class of AGN. We focused on the group of 18 sources that include SAGE0536AGN and SAGE0534AGN, see red box in Figure \ref{fig:tsnemap} (right), that are also separate from the large clusters of sources, in order to find more such objects. A list of these sources, with their identifiers and co-ordinates, can be seen in Table \ref{tab:tsnesample}. The other objects in this t-SNE map are to be explored in a following paper that makes use of more than one machine learning technique and a wider range of multi-wavelength data, which will classify the sources in the direction of the SMC and LMC, as well as estimate the redshifts of extragalactic sources behind the clouds (Pennock et al., {\it in prep.}).

\begin{table}
\caption{Sample of similar sources identified through a t-SNE analysis.}
    \centering
    \begin{tabular}{@{}l@{}l@{}c@{}c@{}}
        \hline\hline 
        Source Nam\rlap{e} & Identifier & RA & DEC  \\
        & & (J2000) & (J2000)\\
       \hline
       SAGE0536AGN\mbox{~} & SAGE1C J053634.78$-$722658.5\mbox{~} & 5:36:34.78 & $-$72:26:58.5  \\
       SAGE0534AGN & SAGE1C J053444.17$-$673750.1 & 5:34:44.17 & $-$67:37:50.1   \\
       1 & WISEA J003617.01$-$743131.3 & 0:36:16.99 & $-$74:31:31.3   \\
       2 & WISEA J011337.10$-$742755.3 & 1:13:37.08 & $-$74:27:55.3   \\
       3 & WISEA J003156.88$-$733113.6 & 0:31:56.89 & $-$73:31:13.6   \\
       4 & WISEA J002602.54$-$724718.0 & 0:26:02.54 & $-$72:47:18.0   \\
       5 & OGLE SMC-LPV-7107 & 0:48:25.71 & $-$72:44:02.8   \\
       6 & WISEA J011408.02$-$723243.1 & 1:14:07.99 & $-$72:32:43.3   \\
       7 & [MCS2008] 11 & 0:55:51.51 & $-$73:31:10.0   \\
       8 & [MA93] 1895 & 1:22:36.94 & $-$73:10:16.7   \\
       9 & MQS J012108.42$-$730713.1 & 1:21:08.43 & $-$73:07:13.1   \\
       10 & MQS J011534.10$-$725049.3 & 1:15:34.09 & $-$72:50:49.3   \\
       11 & WISEA J003910.76$-$713409.9 & 0:39:10.78 & $-$71:34:09.9  \\
       12 & [VV2006] J005116.9$-$721651 & 0:51:16.95 & $-$72:16:51.5  \\
       13 & 2E 238 & 0:57:32.75 & $-$72:13:02.3   \\
       14 & WISEA J013604.46$-$721315.3 & 1:36:04.46 & $-$72:13:15.4   \\
       15 & NAME SMC B0031$-$7042 & 0:34:05.26 & $-$70:25:52.3  \\
       16 & WISEA J004952.56$-$692956.4 & 0:49:52.53 & $-$69:29:56.4   \\
       \hline
    \end{tabular}
    
    \label{tab:tsnesample}
\end{table}

\section{Results}\label{resultssection}

The sample (including SAGE0536AGN and SAGE0534AGN) is made up of 18 sources, 16 of which are spectroscopically confirmed extragalactic sources (see next subsection). Source 8 has no current spectroscopic confirmation of it being an extragalactic source, but has been previously identified as a potential H$\alpha$ emission line star \citep[][identified an emisson line and no underlying continuum]{1993A&AS..102..451M}, a far-infrared (far-IR) object \citep{2011AJ....142..103B,2016MNRAS.457.2814S} and an emission line object \citep{2020A&A...636A..48G}.

Source 5 is a near-superposition of a carbon star in front of the true extragalactic radio source, which can be seen from the spectrum and the annotated spectral lines \citep{1996ApJS..105..419B,1998A&A...329..169V} that is included in the online appendix. 

\subsection{VMC light curves}\label{lightcurvesection}

\begin{figure*} 
\centering
\begin{tabular}{c}
	\includegraphics[trim={0cm 0cm 0cm 1cm},width=\textwidth]{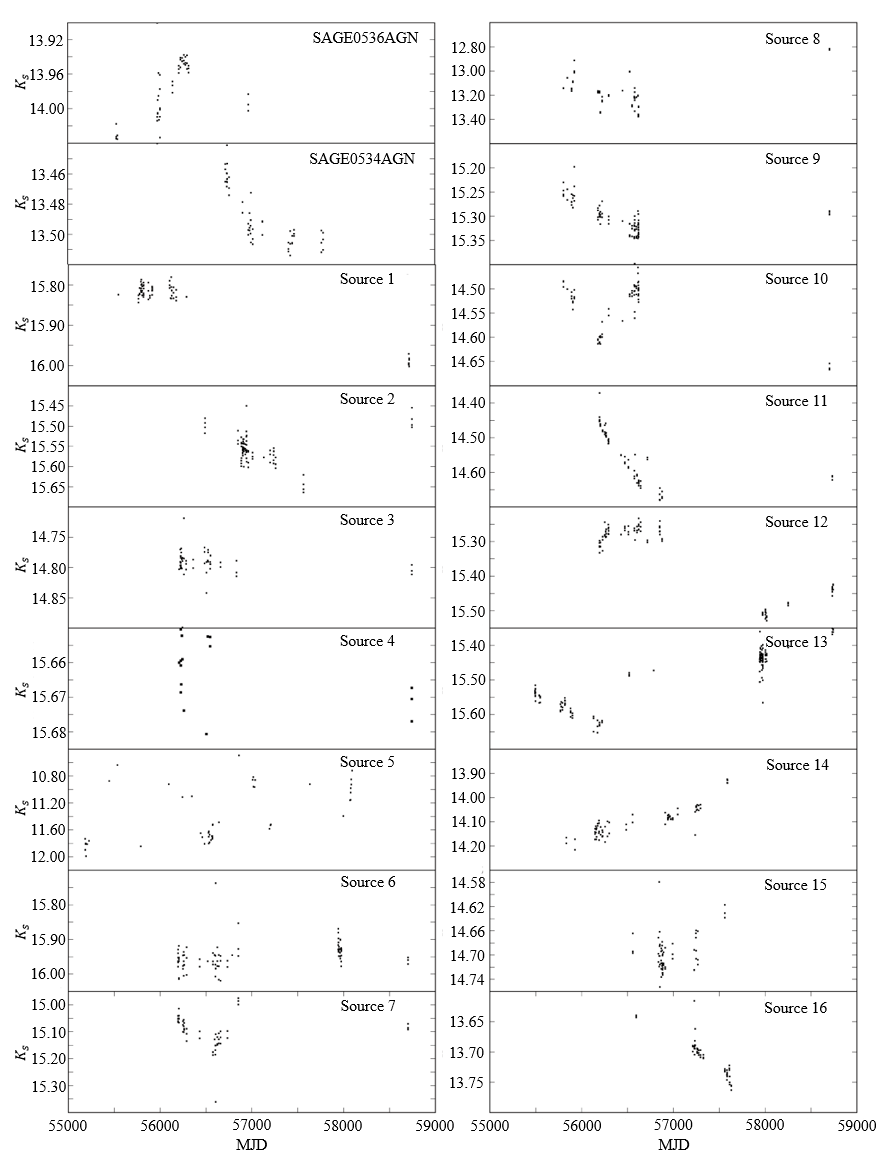}
	\end{tabular}
    \caption{Light curves from the VMC survey of all sources in our t-SNE selected sample.}
    \label{fig:lightcurves}
\end{figure*}

The VMC survey is comprised of multi-epoch observations, which allows for the detection of variability. The light curves of the sample can be seen in Figure \ref{fig:lightcurves}. The amplitudes of variation in $K$\textsubscript{s} were calculated by selecting the highest and lowest set of points. At these points of time the median value is selected as the highest/lowest value. The amplitude is calculated as the difference between these values. For some of these sources we are not seeing the full amplitude, such as for Source 14, where the source becomes brighter without reaching a noticeable peak. The amplitudes calculated from the VMC light curves in $K$\textsubscript{s} can be seen in Table \ref{tab:var}.

\begin{table}
\caption{t-SNE selected sample variability amplitudes and mean magnitudes in the $K$\textsubscript{s} band.}
    \centering
    \begin{tabular}{|l|c|c|}
        \hline\hline 
        Source Name & $K$\textsubscript{s} Amplitude (Vega mag) & Mean $K$\textsubscript{s} (Vega mag)   \\
         &  &   \\
       \hline
       \llap{S}AGE0536A\rlap{GN} & 0.09$\pm$0.01 & 13.49$\pm$0.01 \\
       \llap{S}AGE0534A\rlap{GN} & 0.05$\pm$0.01 & 13.98$\pm$0.03 \\
        1 & 0.18$\pm$0.02 & 15.83$\pm$0.05 \\
        2 & 0.16$\pm$0.02 & 15.34$\pm$0.05 \\
        3 & 0.02$\pm$0.02 & 14.79$\pm$0.02 \\
        4 & 0.02$\pm$0.01 & 15.65$\pm$0.03 \\
        5 & 1.20$\pm$0.13 & 11.45$\pm$0.36 \\
        6 & 0.03$\pm$0.03 & 15.95$\pm$0.04 \\
        7 & 0.16$\pm$0.03 & 15.10$\pm$0.06 \\
        8 & 0.55$\pm$0.07 & 13.19$\pm$0.14 \\
        9 & 0.09$\pm$0.02 & 15.30$\pm$0.03 \\
       10 & 0.17$\pm$0.02 & 14,53$\pm$0.05 \\
       11 & 0.22$\pm$0.01 & 14.56$\pm$0.07 \\
       12 & 0.25$\pm$0.01 & 15.34$\pm$0.10 \\
       13 & 0.27$\pm$0.01 & 15.49$\pm$0.08 \\
       14 & 0.26$\pm$0.02 & 14.13$\pm$0.32 \\
       15 & 0.09$\pm$0.02 & 14.70$\pm$0.03 \\
       16 & 0.12$\pm$0.01 & 13.70$\pm$0.08 \\
       \hline
    \end{tabular}
    
    \label{tab:var}
\end{table}

Source 5 shows large-amplitude, semi-regular variability that corroborates its identity as a carbon star. The variability of the other sources combined with their extragalactic spectroscopic confirmation, confirms the presence of an AGN. However, Sources 3, 4 and 6 show little to no variability.

\subsection{Optical line identifications and spectral analysis}\label{spectrasection}

\begin{figure*}
\centering
\begin{tabular}{c}
	\includegraphics[trim={0.5cm 0.5cm 0cm 0cm},width=1\textwidth]{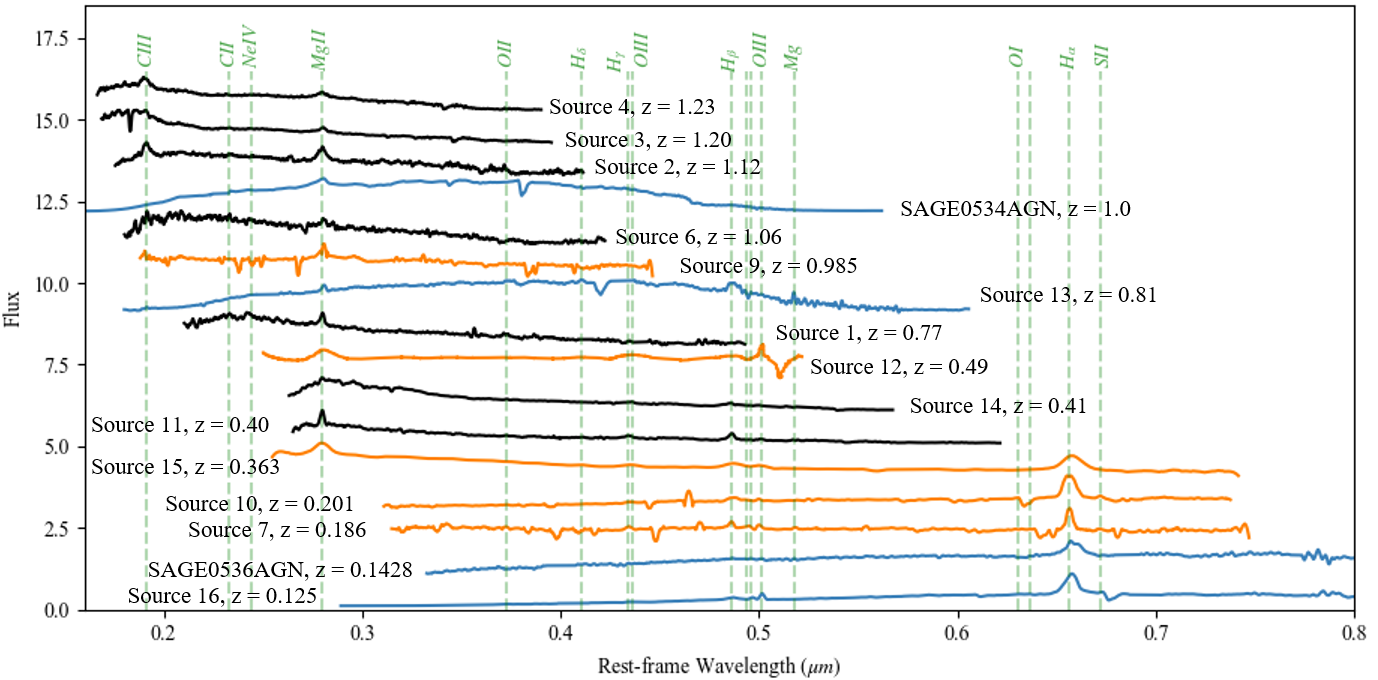}
	\end{tabular}
    \caption{Spectra of t-SNE selected sample, excluding Source 5 (carbon star, spectrum shown in online appendix) and 8 (no spectrum available). Spectral flux has been normalised. Black indicates sources that were observed with SAAO 1.9m telescope and blue indicates sources observed with SALT. Orange indicates sources that were observed prior to this study with other facilities.}
    \label{fig:spectra}
\end{figure*}

The spectra of SAGE0536AGN, SAGE0534AGN and 14 out of 16 t-SNE sample sources that were observed with SALT, SAAO's 1.9m telescope or with other facilities prior to this study are shown in Figure \ref{fig:spectra}. Sources that were observed as part of other surveys are also shown in Figure \ref{fig:spectra}. Only Sources 5 (Star) and 8 (no available spectrum) are not shown. The spectra of Source 5 and sources with multiple available spectra that are not shown in Figure \ref{fig:spectra} can be found in the online appendix. Redshifts are listed in Table \ref{tab:knownAGNclass}. The redshift for Source 8 was estimated from photometry \citep{2015PASA...32...10F,2021yCat.7290....0F}.

The Full Width at Half Maximum (FWHM) is calculated by modelling the continuum surrounding the emission line and then subtracting the continuum from the spectra. After this the half maximum height of the emission line is calculated from the line profile and then subsequently the width of the emission to get the observed FWHM. The intrinsic FWHM is then calculated from $FWHM_{\rm intrinsic}=\sqrt{(FWHM_{\rm observed})^{2}-(FWHM_{\rm instrument})^{2}}$, where $FWHM_{\rm instrument}$ is the FWHM of the instrument used to obtain the spectrum.

\begin{table*}
\caption{Table of sources investigated in this work. Redshifts are calculated from spectroscopy, except for source 8 (indicated with an *) for which the redshift was calculated from photometry. (1) \protect\cite{2015MNRAS.453.2341V}; (2) \protect\cite{2013ApJ...775...92K}; (3) \protect\cite{2015PASA...32...10F, 2021yCat.7290....0F}; (4) \protect\cite{2003AJ....125....1G}; (5) Kishimoto et al. (\textit{in prep.}). Source 5 was found to be a carbon star in the SMC dominating in the optical/IR. }
    \centering
    \begin{tabular}{|l|c|c|c|ccccc}
        \hline\hline 
        Source Name & $z$ & Emission Lines & \multicolumn{3}{c}{--- FWHM (km s\textsuperscript{-1}) ---} & Date &  Ref. \\
         &  &  & H$\alpha$ & H$\beta$ & Mg\,{\sc ii} &   Observed &   \\
       \hline 
       SAGE0536AGN  & 0.1428$\pm$0.0001 & H$\alpha$ & \mbox{~~}3900$\pm$450 & & & SALT \mbox{~}08-09-2012 & (1)\\
       SAGE0534AGN  & 1.0$\pm$0.01 & Mg\,{\sc ii} & &  & 6450$\pm$200 &  SAAO 16-11-2019 & This work. \\
       SAGE0534AGN  & 1.009$\pm$0.002 & Mg\,{\sc ii} & &  & 10310$\pm$300  &  SALT \mbox{~}01-11-2021/ & This work. \\
        & & & & & & \mbox{~}\mbox{~}\mbox{~}\mbox{~}\mbox{~}\mbox{~}\mbox{~}\mbox{~}\mbox{~}\mbox{~}\mbox{~}17-03-2022 & \\
       1  & 0.77$\pm$0.01 & Mg\,{\sc ii} & &  & 2300$\pm$250 &  SAAO 24-11-2019 & This work. \\
       2  & 1.12$\pm$0.01 & C\textsc{III}, Mg\,{\sc ii} & &  & 5100$\pm$550 & SAAO 31-10-2019 & This work. \\
       3  & 1.2$\pm$0.01 & Mg\,{\sc ii} & &  & 3800$\pm$350 & SAAO 05-11-2019 & This work. \\
       4  & 1.23$\pm$0.02 & C\textsc{III}, Mg\,{\sc ii} &   & & 6100$\pm$200 & SAAO 22-11-2019 & This work. \\
       5 & -- & & & & &  SALT \mbox{~}17-07-2021 & This work.\\
       6  & 1.06$\pm$0.02 & Mg\,{\sc ii} & & & & SAAO 29-10-2019 &  This work. \\
       7 & 0.186$\pm$0.005 & H$\alpha$, H$\beta$ & && & 02-2012 -- 01-2013 &   (2) \\
       8 & 0.5* & & & & &  N/A &  (3) \\
       9 & 0.985$\pm$0.005 & Mg\,{\sc ii} & & & 3350$\pm$500 & 02-2012 -- 01-2013 &   (2) \\
       10 & 0.201$\pm$0.005 & H$\alpha$, H$\beta$ & 3050$\pm$1000 & 3700$\pm$1000 & & 02-2012 -- 01-2013 &  (2) \\
       11  & 0.4$\pm$0.01 & H$\beta$, H$\gamma$ and Mg\,{\sc ii} & & 1900$\pm$250 & 2200$\pm$250 & SAAO 02-11-2019 &  This work. \\
       12 & 0.49$\pm$0.005 & Mg\,{\sc ii}, H$\gamma$, H$\beta$, OIII  &  & & 5750$\pm$500 & 10-1999 -- 01-2001 &  (4) \\
       13  & 0.81$\pm$0.02 & Mg\,{\sc ii}, H$\beta$ & & &  & SAAO 19-11-2019 &  This work. \\
       13  & 0.81$\pm$0.02 & Mg\,{\sc ii}, H$\beta$ & & & 6300$\pm$350 & SALT \mbox{~}24-07-2021 &  This work. \\
       14  & 0.41$\pm$0.01 & Mg\,{\sc ii}, H$\beta$ & & &  & SAAO 30-10-2019 &  This work. \\
       15 & 0.363$\pm$0.005 & Mg\,{\sc ii}, H$\gamma$, H$\beta$, OIII, H$\alpha$  & 5450$\pm$900 & 6150$\pm$300 & 6450$\pm$500 &  \mbox{~~}\mbox{~~}\mbox{~~}\mbox{~~}\mbox{~~}\mbox{~~}22-08-2004 &  (5) \\
       16 & 0.125$\pm$0.01 & H$\alpha$, H$\beta$, OIII & 2050$\pm$450 & 2800$\pm$300 &  & SALT \mbox{~}01-09-2021 &  This work \\
       \hline
    \end{tabular}
    \label{tab:knownAGNclass}
\end{table*}

The continuum of the optical spectra was modelled using Python's Astropy module \citep{2013A&A...558A..33A,2018AJ....156..123A}. This facilitated the use of a low order polynomial to model the continuum. The most notable emission line is the Mg\,{\sc ii} $\lambda 2798$ line, observed in all but four of the sources.

\subsubsection{Black hole masses}

The sample showcases emission lines from either Mg\,{\sc ii} $\lambda 2798$, H$\alpha$ $\lambda 6563$ \AA\ or H$\beta$ $\lambda 4861$ \AA. From these we can calculate the black hole masses ($M$\textsubscript{BH}). The black hole mass was not calculated for Sources 6 and 14 due to their spectra being too noisy, and Sources 1, 7 and 11 had FWHM that were too close to the instrumental FWHM to be disentangled.

The calibrations used for calculating the black hole masses from the Mg\,{\sc ii} \citep{2012MNRAS.427.3081T}, H$\beta$ \citep{2006ApJ...641..689V} and H$\alpha$ \citep{2019MNRAS.487.3404B} emission lines are:

\begin{equation} \label{eq:MgII}
    M_{\rm BH}=10^{6.748} \left( \frac{L_{3000}}{10^{44} \rm erg/s}\right)^{0.620} \left( \frac{FWHM_{\rm Mg\,II}}{10^3 \rm km/s} \right)^2 M_{\odot}
\end{equation}

\begin{equation} \label{eq:Hb}
    M_{\rm BH}=10^{6.910} \left( \frac{L_{5100}}{10^{44} \rm erg/s}\right)^{0.500} \left( \frac{FWHM_{\rm H\beta}}{10^3 \rm km/s} \right)^2 M_{\odot}
\end{equation}

\begin{equation} \label{eq:Ha}
    M_{\rm BH}=\epsilon 10^{6.9} \left( \frac{L_{5100}}{10^{44} \rm erg/s}\right)^{0.54} \left( \frac{FWHM_{\rm H\alpha}}{10^3 \rm km/s} \right)^{2.06} M_{\odot}
\end{equation}

where $L$\textsubscript{3000} and $L$\textsubscript{5100} are the monochromatic continuum luminosities at rest-frame 3000 \AA\ and 5100 \AA\ respectively in erg s\textsuperscript{-1} derived from SED fitting (see Section \ref{cigalesection} for more information). Calculated black hole masses are shown in Table \ref{tab:BHmasses}. The constant $\epsilon = 1.075$ \citep{2015ApJ...813...82R} was adopted. Error on the monochromatic continuum luminosities were calculated by the SED fitter, errors on the FWHM is the standard error of the mean measurement from the line profile and error of the black hole masses is propagated from these two errors.  

\begin{table*}
\caption{Black hole masses calculated using equations \ref{eq:MgII}, \ref{eq:Hb} and \ref{eq:Ha}. $L$\textsubscript{bol}(AGN) is the AGN bolometric luminosity calculated during SED fitting (see Section \ref{cigalesection}). Eddington Ratio is defined as the $L$\textsubscript{bol}(AGN)/$L$\textsubscript{edd}, where the $L$\textsubscript{edd}=1.25$\times10^{38}$$M$\textsubscript{BH}erg s\textsuperscript{-1}. Sources that are not listed here either have a noisy spectrum or emission lines that are smaller or close to the FWHM of the instrument.}
    \centering
    \begin{tabular}{|l|c|c|c|cc}
        \hline\hline 
        Source Name & \multicolumn{3}{c}{--- $M$\textsubscript{BH} ($M$\textsubscript{$\odot$})---} & $L$\textsubscript{bol}(AGN)  &  Edd. Ratio \\
         & H$\alpha$ & H$\beta$ & Mg\,{\sc ii} & (erg s\textsuperscript{-1}) & ($\%$)   \\
       \hline 
       SAGE0536AGN  & (5.5$\pm$1.3)$\times10^{7}$& - & - & (4.4$\pm$0.2)$\times10^{44}$ & \mbox{~~}6.3$\pm$\mbox{~~}1.5 \\
       SAGE0534AGN  & - & - & (1.9$\pm$0.1)$\times10^{10}$ & (2.4$\pm$0.1)$\times10^{47}$ & \mbox{~~}9.8$\pm$\mbox{~~}0.8\\
       2  & - & - & (2.9$\pm$0.6)$\times10^{9}$ & (9.5$\pm$0.5)$\times10^{46}$ & 26.0$\pm$\mbox{~~}5.6\\
       3  & - & - & (1.8$\pm$0.3)$\times10^{9}$ & (1.9$\pm$0.1)$\times10^{47}$ & 80.4$\pm$15.5\\
       4  & - & - & (3.3$\pm$0.2)$\times10^{9}$ & (7.8$\pm$0.4)$\times10^{46}$ & 18.6$\pm$\mbox{~~}1.6\\
       9  & - & - & (6.9$\pm$2.1)$\times10^{8}$ & (4.5$\pm$0.2)$\times10^{46}$ & 52.3$\pm$15.9\\
       10 & (6.7$\pm$4.6)$\times10^{7}$ & (9.2$\pm$4.7)$\times10^{7}$ & - & (6.4$\pm$0.3)$\times10^{44}$ & \mbox{~~}6.4$\pm$\mbox{~~}3.9\\
       12 & - & - & (2.9$\pm$0.5)$\times10^{8}$ & (5.1$\pm$0.3)$\times10^{45}$ & 14.2$\pm$\mbox{~~}2.6\\
       13  & - & - & (1.2$\pm$0.1)$\times10^{9}$ & (2.8$\pm$0.2)$\times10^{46}$ & 18.0$\pm$\mbox{~~}2.6\\
       15 & (5.7$\pm$1.9)$\times10^{8}$ & (5.0$\pm$0.5)$\times10^{8}$ & - & (6.5$\pm$0.3)$\times10^{45}$ & \mbox{~~}9.7$\pm$\mbox{~~}3.7\\
       16 & (1.7$\pm$0.8)$\times10^{7}$ & (3.1$\pm$0.7)$\times10^{7}$ & - & (5.7$\pm$0.3)$\times10^{44}$ & 19.1$\pm$\mbox{~~}8.2\\
       \hline
    \end{tabular}
    
    \label{tab:BHmasses}
\end{table*}

SAGE0536AGN's black hole mass was previously reported as $M$\textsubscript{BH} $= (3.5 \pm 0.8) \times 10^{8} $M\textsubscript{$\odot$}, and $L$\textsubscript{bol} $= (5.5 \pm 1.3) \times 10^{45}$ erg s\textsuperscript{-1} ($\approx$ 12 $\%$ of the Eddington luminosity \citep{2015MNRAS.453.2341V}. In this work the black hole mass of SAGE0536AGN is calculated from the H$\alpha$ line to be M$_{BH}$=(5.5$\pm$1.3)$\times10^{7}$M$_{\odot}$ with an Eddington ratio of $\sim$6 $\%$. This mass combined with the calculated velocity dispersion, $s\sim$ 202 $\pm$ 15 km s$^{-1}$, puts SAGE0536AGN in agreement with the known correlation between velocity dispersion and black hole mass \citep[e.g.][]{graham_2008}.


\subsection{C\textsc{igale} modelling}\label{cigalesection}
\subsubsection{The code and models}

Code Investigating GALaxy Emission \citep[C\textsc{igale};][]{2009A&A...507.1793N,2019A&A...622A.103B,2020MNRAS.491..740Y,2022arXiv220103718Y}, is a versatile Python code for studying the evolution of galaxies by modelling the X-ray to radio spectrum of galaxies and estimating their physical properties such as star formation rate, attenuation, dust luminosity, stellar mass and characteristics of an active nucleus. It does this by comparing modelled galaxy SEDs to observed ones.

The AGN model of C\textsc{igale} is from \cite{2006MNRAS.366..767F} and assumes that the dusty torus is a smooth structure. However, more recent theoretical and observational works find that the torus is mainly made of dusty clumps \citep[e.g.][]{2009ApJ...707.1550N,2012MNRAS.420.2756S}. Recently, \cite{2022arXiv220103718Y} developed an updated version of C\textsc{igale}, which allows for the modelling of the X-ray emission to account for X-ray fluxes in the fits of the SED. This version also includes a more recent AGN model, with a clumpy two-phase torus model derived from a radiative-transfer method \citep[SKIRTOR model;][]{2012MNRAS.420.2756S,2016MNRAS.458.2288S}. This model also accounts for the presence of AGN polar dust extinction that has been observed in type 1 AGN \citep{2015torus}. Furthermore, the radio models now account for radio emission from an AGN, not just star formation as it did previously. It is this version of the code that we use in this work.

The SKIRTOR model is a library of AGN dusty torus emission models that were calculated with SKIRT, a radiative transfer code based on a Monte Carlo technique. In this model the dust distribution of the torus is modelled as a two phase medium. This medium consists of a large number of high-density clumps embedded in a smooth dusty component of low density. The advantage of this model is that it can produce both  attenuated silicate features and pronounced near-IR emission at the same time, which both smooth and clumpy models find challenging. Since SKIRTOR's creation evidence, both simulated  \citep{2013MNRAS.429.1494R} and observational \citep{2013ASSP...34..331P,2014MNRAS.439.1403M,2015ApJ...809L..13L}, have shown that the dusty torus is a multi-phase structure.

The C\textsc{igale} fit is made of a maximum of eight modules. The first is the star formation history (SFH) module, the one used here is the delayed SFH with optional exponential burst which 
provides efficient modelling of early–type and late–type galaxies. The second is the simple stellar population module that computes the intrinsic stellar spectrum, for which we selected the standard \cite{2003MNRAS.344.1000B} model. The  modified dust attenuation law from \cite{2000ApJ...533..682C} is our third module, which controls the UV attenuation with the colour excess E(B$-$V), and also the power-law slope ($\delta$) that modifies the attenuation curve. We included the nebular emission module, though we kept default parameters. The module to model the dust emission in the SED uses a modified blackbody spectrum following \cite{2014ApJ...784...83D}. Next is the AGN module, modelled as a two phase torus \citep{2012MNRAS.420.2756S,2016MNRAS.458.2288S}, where we set the extinction law of the polar dust to the SMC values \citep{1984A&A...132..389P}, the temperature to 100 K \citep[e.g.][]{2021A&A...654A..93B} and the emissivity index of the polar dust to 1.6 \citep{2012MNRAS.425.3094C}. The radio module is also included as all the sources have radio observations, as the recent update to C\textsc{igale} \citep{2022arXiv220103718Y} now models radio emission from an AGN. Where there is only one radio observation, the spectral index, $\alpha$, is set to the default of $-$0.7, typical of synchrotron emission. Where the sources have X-ray observations the X-ray module was implemented.

\subsubsection{Inputs}

The known redshifts and photometry from SMASH, Gaia EDR3, VMC, SAGE, AllWISE and HERITAGE \citep{2013HERITAGE} were used to model the SEDs of the 17 objects. Not all sources had far-IR fluxes, due to either being outside of the HERITAGE survey field of the Magellanic Clouds or the fluxes being too faint. Where far-IR fluxes were not found in images an upper limit on the flux was calculated from the HERITAGE images. For Source 8, where there was no spectroscopically determined redshift, the photometric redshift, calculated by \cite{2015PASA...32...10F, 2021yCat.7290....0F} was used.

Models used and the parameters that were varied over the fit are shown in Table \ref{CIGALEmodels}. Each AGN was initially fit without extragalactic dust model and where the models did not fit in the far-IR and showed $f$\textsubscript{AGN}$ < 0.99$, where $f$\textsubscript{AGN} is the fraction of the total dust that is due to the AGN, the extragalactic dust model was then added, which is the case for five of the AGN.

\begin{table*}
	\centering
	\caption{Modules and parameter values used to model the sample in C\textsc{igale}. For the parameter values not listed the default values were used.} 
	\begin{tabular}{l|l|l}  
		\hline\hline
        Parameter & Model/Values & Description \\
		\hline\hline
		 Star formation history (SFH) & delayed SFH with optional exponential burst &\\
		\hline
		$\tau_{\rm main}$ & 100 -- 4000 & e-folding time of main stellar population model (Myr).\\
		$t$ & 100 -- 6000 & Age of oldest stars in the galaxy (Myr).\\
		\hline \hline
		 Simple stellar population (SSP) & \cite{2003MNRAS.344.1000B} \\
		\hline
		IMF & 0 & Initial Mass Function from \cite{2003PASP..115..763C}\\
		Metallicity & 0.0001, 0.01, 0.02, 0.05 & Metallicity, where solar metallicity $\sim$ 0.02.\\
		Separation age & 1, 5, 10 & Separation between young and old star populations (Myr).\\
		\hline\hline
		 Galactic dust attenuation & Modified \cite{2000ApJ...533..682C} attenuation law & \\
		\hline
		$E(B - V)$ & 0.4 & Colour excess of nebular lines (mag).\\
		Ext\textunderscore law \textunderscore emission \textunderscore lines & LMC, SMC & Extinction law for attenuating emission lines flux \citep{1992ApJ...395..130P}.\\
		\hline\hline
		Galactic dust emission & \cite{2014ApJ...784...83D}\\
		\hline
		$\beta$ & 0.0625 -- 4 & Slope in $dM_{\rm dust} \propto U^{-\beta}dU$\\
		\hline\hline
		 AGN & SKIRTOR UV-to-IR,  from \cite{2012MNRAS.420.2756S,2016MNRAS.458.2288S} &\\
        \hline
        $\tau$ &  3, 5, 7, 9, 11 & Optical depth at 9.7 $\mu$m.\\
        $pl$ & 0, 0.5, 1, 1.5 & Torus radial density parameter, such that \\
        & & $\rho \propto r^{-pl}e^{-q|cos(\theta)|}$, where $\rho$ is the torus density and $r$ \\
        & & is the radius of the torus.\\
        $q$ & 0, 0.5, 1, 1.5 & Torus density angular parameter.\\
        Opening Angle & 10, 20, 30, 40, 50, 60, 70, 80 &  Angle between the equatorial plane and edge of the torus.\\
        $R$ & 10, 20, 30 & Ratio of the outer to inner radii of the dust torus, $R_{\rm out}/R_{\rm in}$\\
        $i$ & 0, 10, 20, 30, 40, 50, 60, 70, 80, 90 & Viewing angle where face-on: $i = 0^\circ$, edge-on: $i = 90^\circ$ \\
        $f_{\rm AGN}$ & 0.6, 0.7, 0.8, 0.9, 0.999 & AGN fraction, $f_{\rm AGN} = \frac{L_{\rm dust,AGN}}{L_{\rm dust,AGN}+L_{\rm dust,galaxy}}$, where \\
         & & $L_{\rm dust,AGN}$ and $L_{\rm dust,galaxy}$ are AGN and galaxy dust \\
         & & luminosity integrated over all wavelengths, respectively. \\
        $\delta$ & $-$0.36 -- 0.36 & Power-law modifying the optical slope of the disk.\\
        Law$_{\rm polar}$ & SMC & Extinction law of polar dust.\\
        $E(B - V)_{\rm polar}$ & 0, 0.05, 0.1, 0.2, 0.3, 0.4, 0.5, 0.6, 0.7, 0.8, 0.9, 1.0 & Polar-dust colour excess (mag).\\
        $T_{\rm polar}$ & 100 K & Temperature of polar dust.\\
        Emissivity$_{\rm polar}$ & 1.6 & Emissivity index of polar dust \\
        & & \citep[see equation (10) of][]{2020MNRAS.491..740Y}.\\
		\hline\hline
		X-ray\\
		\hline
		\textit{$\Gamma$} & 1.5 -- 2.0 & Photon index, $\Gamma$, of the AGN intrinsic X-ray spectrum.\\
		$\alpha_{ox}$ & $-$1.9, $-$1.8, $-$1.7, $-$1.6, $-$1.5, $-$1.4 & UV/X-ray slope calculated at $i$ = 30\textdegree.\\
		\hline\hline
		Radio\\
		\hline
		$\alpha_{\rm SF}$ & 0.8 & Slope of the power-law synchrotron emission related to SF, \\
		& & which is a free power-law slope.\\
		$\alpha_{\rm AGN}$ & 0.01 -- 2 & Slope of the power-law AGN radio emission, defined as \\
		& & $L_{\rm \nu, AGN} \propto \nu^{\rm -\alpha_{AGN}}$.\\
		$R_{\rm AGN}$ & 0.1 -- 300 & Radio-loudness parameter, defined as $L_{\rm \nu, 5GHz}/L_{\rm \nu, 2500\AA}$,\\
		& &  where $L_{\rm \nu,5GHz}$ and $L_{\rm \nu, 2500\AA}$ are the monochromatic AGN \\
		& &  luminosities per frequency at rest-frame 5 GHz and 2500 \AA. \\
		\hline
	\end{tabular}
	\label{CIGALEmodels}
\end{table*}

\subsubsection{C\textsc{igale} models for SAGE0536AGN and SAGE0534AGN}

\begin{figure*}
\centering
\begin{tabular}{c}
	\includegraphics[trim={0.5cm 1cm 0cm 0.7cm},width=1\textwidth]{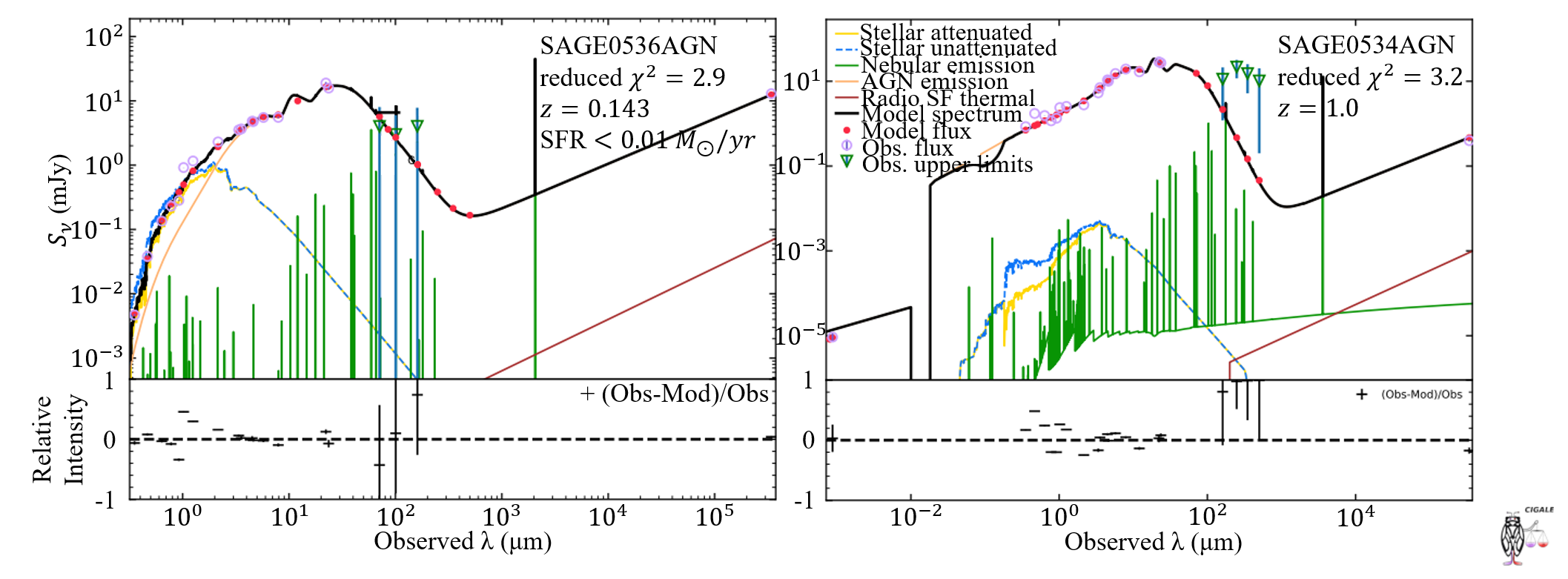}
	\end{tabular}
    \caption{C\textsc{igale} best fits of SAGE0536AGN (left) and SAGE0534AGN (right). SED fits of the rest of the sample can be found in the Appendix.}
    \label{fig:XCIGALE1}
\end{figure*}

C\textsc{igale} SED fits of SAGE0534AGN and SAGE0536AGN are shown in Figure \ref{fig:XCIGALE1} and the calculated parameters can be found in Table \ref{CIGALEmodelsout}.

\begin{table*}
	\centering
	\caption{AGN properties calculated with C\textsc{igale}. AGN fraction is the fraction of IR luminosity from the object that is due to the AGN. $\tau$ is the torus optical depth at 9.7 $\mu$m. The inclination angle, $i$, is the viewing angle, where $i = 0^\circ$ is face-on and $i = 90^\circ$ is edge-on. $R$ is the ratio between the maximum and minimum radii of the torus. The opening angle, $oa$, is the angle between the equatorial plane and edge of the torus. $pl$ is the torus radial density parameter and $q$ is the torus density angular parameter, such that $\rho \propto r^{-pl}e^{-q|cos(\theta)|}$, where $\rho$ is the torus density and $r$ is the radius of the torus. $E(B-V)$ is the extinction caused by polar dust. Accretion power is the intrinsic AGN disk luminosity averaged over all directions. AGN luminosity is the sum of the observed AGN disk luminosity (some might be extincted) and the observed AGN dust re-emitted luminosity. AGN torus fraction is the fraction of the AGN luminosity that is re-emitted by the torus dust.} 
	\begin{tabular}{lccccccccccc}  
		\hline\hline
        Source & AGN & $\tau$ & $i$ & $R$ &  $oa$  & $pl$ & $q$ &  $E(B-V)$  & Accretion  & AGN & Torus \\
        & fraction & & \llap{(}degrees\rlap{)} & & \llap{(}degrees\rlap{)} &  &  &   &  power &   luminosity  & fraction \\
        & & &  & &  &  &  &   &  (10$^{37}$W) &  (10$^{37}$W) &  \\
		\hline
		\llap{S}AGE0536AG\rlap{N} & 0.90$\pm$0.0\rlap{1} & 3.0$\pm$0.\rlap{1} & \mbox{~~}3.4$\pm$\mbox{~~}4.\rlap{7}  & 29.1$\pm$2.\rlap{9} & 79.5$\pm$2.\rlap{2} & 0.5$\pm$0.\rlap{2} & 1.2$\pm$0.\rlap{3} & 0.96$\pm$0.0\rlap{1}  & \mbox{~~}\mbox{~~}2.7$\pm$\mbox{~~}0.\rlap{2} & \mbox{~~}\mbox{~~}\mbox{~~}4.4$\pm$\mbox{~~}\mbox{~~}0.\rlap{2} & 0.72$\pm$0.05 \\
		\llap{S}AGE0534AG\rlap{N} & 0.99$\pm$0.0\rlap{1} & 4.0$\pm$1.\rlap{0} & \mbox{~~}4.5$\pm$\mbox{~~}4.\rlap{9}  & 26.2$\pm$4.\rlap{9} & 40.1$\pm$1.\rlap{0} & 1.2$\pm$0.\rlap{3} & 1.0$\pm$0.\rlap{4} & 0.00$\pm$0.0\rlap{1}  & 772.1$\pm$38.\rlap{6} & 2375.9$\pm$118.\rlap{8} & 0.17$\pm$0.01 \\
		Source 1 & 0.98$\pm$0.0\rlap{2} & 7.4$\pm$1.\rlap{5} & 10.3$\pm$\mbox{~~}7.\rlap{9}   & 23.0$\pm$6.\rlap{5} & 65.9$\pm$7.\rlap{7} & 1.2$\pm$0.\rlap{2} & 0.2$\pm$0.\rlap{2} & 0.10$\pm$0.0\rlap{1} &  \mbox{~~}69.4$\pm$\mbox{~~}6.\rlap{9} & \mbox{~~}171.4$\pm$\mbox{~~}\mbox{~~}8.\rlap{6} & 0.52$\pm$0.04 \\
		Source 2 & 0.99$\pm$0.0\rlap{1} & 4.1$\pm$1.\rlap{0} & 21.9$\pm$\mbox{~~}9.\rlap{5}  & 16.6$\pm$7.\rlap{5} & 40.3$\pm$1.\rlap{6} & 0.6$\pm$0.\rlap{3} & 1.2$\pm$0.\rlap{2} & 0.00$\pm$0.0\rlap{1} &  358.1$\pm$29.\rlap{2} & \mbox{~~}953.6$\pm$\mbox{~~}47.\rlap{7} & 0.17$\pm$0.02 \\
		Source 3 & 0.91$\pm$0.0\rlap{1} & 3.3$\pm$0.\rlap{7} & \mbox{~~}5.7$\pm$\mbox{~~}5.\rlap{0}   & 29.0$\pm$3.\rlap{1} & 50.0$\pm$0.\rlap{1} & 1.4$\pm$0.\rlap{2} & 1.3$\pm$0.\rlap{2} & 0.04$\pm$0.0\rlap{1} &  738.9$\pm$36.\rlap{9} & 1850.2$\pm$\mbox{~~}92.\rlap{5} & 0.30$\pm$0.02\\
		Source 4 & 0.99$\pm$0.0\rlap{1} & 4.7$\pm$1.\rlap{6} & 10.7$\pm$\mbox{~~}9.\rlap{0}  & 27.1$\pm$4.\rlap{6} & 54.9$\pm$5.\rlap{0} & 1.2$\pm$0.\rlap{3} & 1.0$\pm$0.\rlap{4} & 0.00$\pm$0.0\rlap{1}  &  233.1$\pm$17.\rlap{4} & \mbox{~~}774.8$\pm$\mbox{~~}38.\rlap{7} & 0.28$\pm$0.03\\
		Source 6 & 0.94$\pm$0.0\rlap{1} & 3.7$\pm$0.\rlap{9} & 24.4$\pm$\mbox{~~}7.\rlap{0}  & 29.0$\pm$3.\rlap{0}  & 53.9$\pm$4.\rlap{9} & 1.2$\pm$0.\rlap{2} & 0.6$\pm$0.\rlap{4} & 0.30$\pm$0.0\rlap{1} &  124.3$\pm$14.\rlap{1} & \mbox{~~}329.7$\pm$\mbox{~~}16.\rlap{5} & 0.33$\pm$0.03 \\
		Source 7 & 0.72$\pm$0.0\rlap{1} & 5.8$\pm$1.\rlap{4} & 49.3$\pm$\mbox{~~}2.\rlap{5}   & 24.8$\pm$6.\rlap{3} & 40.7$\pm$2.\rlap{5} & 0.9$\pm$0.\rlap{2} & 1.3$\pm$0.\rlap{2} & 0.22$\pm$0.0\rlap{3} &  \mbox{~~}\mbox{~~}5.7$\pm$\mbox{~~}0.\rlap{3} & \mbox{~~}\mbox{~~}\mbox{~~}5.9$\pm$\mbox{~~}\mbox{~~}0.\rlap{3} & 0.43$\pm$0.04 \\
		Source 8 & 0.89$\pm$0.0\rlap{1} & 4.6$\pm$0.\rlap{8} & 21.8$\pm$\mbox{~~}3.\rlap{9}   & 28.1$\pm$3.\rlap{9} & 68.2$\pm$3.\rlap{9} & 1.4$\pm$0.\rlap{2}  & 1.5$\pm$0.\rlap{1} & 0.08$\pm$0.0\rlap{1} &  117.6$\pm$12.\rlap{4} & \mbox{~~}223.1$\pm$\mbox{~~}11.\rlap{2}  & 0.58$\pm$0.04 \\
		Source 9 & 0.98$\pm$0.0\rlap{2} & 3.8$\pm$1.\rlap{0} & \mbox{~~}4.6$\pm$\mbox{~~}5.\rlap{0}  & 26.5$\pm$4.\rlap{8} & 53.3$\pm$4.\rlap{7} & 0.4$\pm$0.\rlap{3} & 1.3$\pm$0.\rlap{2} & 0.00$\pm$0.0\rlap{1} &  136.5$\pm$\mbox{~~}6.\rlap{8} & \mbox{~~}447.7$\pm$\mbox{~~}22.\rlap{4}  & 0.24$\pm$0.02 \\
		Source 10 & 0.70$\pm$0.0\rlap{2} & \llap{1}0.1$\pm$1.\rlap{3} & \mbox{~~}2.5$\pm$\mbox{~~}4.\rlap{5}  & 29.4$\pm$2.\rlap{4} & 74.0$\pm$6.\rlap{0} & 0.6$\pm$0.\rlap{3} & 0.5$\pm$0.\rlap{4} & 0.30$\pm$0.0\rlap{1} &  \mbox{~~}\mbox{~~}2.7$\pm$\mbox{~~}0.\rlap{1} & \mbox{~~}\mbox{~~}\mbox{~~}6.4$\pm$\mbox{~~}\mbox{~~}0.\rlap{3} & 0.65$\pm$0.05 \\
		Source 11 & 0.96$\pm$0.0\rlap{3} & 5.4$\pm$1.\rlap{9} & 21.8$\pm$\mbox{~~}9.\rlap{6}  & 28.1$\pm$3.\rlap{9} & 52.6$\pm$4.\rlap{5} & 1.2$\pm$0.\rlap{2} & 1.1$\pm$0.\rlap{4} & 0.08$\pm$0.0\rlap{1} &  \mbox{~~}38.3$\pm$\mbox{~~}4.\rlap{9} & \mbox{~~}\mbox{~~}82.6$\pm$\mbox{~~}\mbox{~~}5.\rlap{2} & 0.41$\pm$0.03\\
		Source 12 & 0.99$\pm$0.0\rlap{1} & 9.1$\pm$1.\rlap{6} & 19.1$\pm$11.\rlap{5}  & 24.7$\pm$6.\rlap{0} & 46.7$\pm$4.\rlap{8} & 0.6$\pm$0.\rlap{3} & 0.4$\pm$0.\rlap{4} & 0.12$\pm$0.0\rlap{2} &  \mbox{~~}24.0$\pm$\mbox{~~}3.\rlap{4} & \mbox{~~}\mbox{~~}50.6$\pm$\mbox{~~}\mbox{~~}2.\rlap{5}  & 0.40$\pm$0.04 \\
		Source 13 & 0.93$\pm$0.0\rlap{3} & 8.5$\pm$1.\rlap{6} & \mbox{~~}5.7$\pm$\mbox{~~}6.\rlap{4}   & 23.8$\pm$6.\rlap{2} & 62.7$\pm$6.\rlap{5} & 0.5$\pm$0.\rlap{2} & 1.0$\pm$0.\rlap{4} & 0.10$\pm$0.0\rlap{2} & 119.3$\pm$13.\rlap{5} & \mbox{~~}276.7$\pm$\mbox{~~}22.\rlap{9} & 0.52$\pm$0.06 \\
		Source 14 & 0.98$\pm$0.0\rlap{2} & 5.7$\pm$1.\rlap{3} & 12.3$\pm$\mbox{~~}9.\rlap{7}  & 25.8$\pm$5.\rlap{5} & 50.6$\pm$2.\rlap{7} & 0.9$\pm$0.\rlap{3} & 0.4$\pm$0.\rlap{4} & 0.00$\pm$0.0\rlap{2} &  \mbox{~~}29.9$\pm$\mbox{~~}2.\rlap{2} & \mbox{~~}\mbox{~~}98.6$\pm$\mbox{~~}\mbox{~~}4.\rlap{9} & 0.28$\pm$0.02 \\
		Source 15 & 0.90$\pm$0.0\rlap{2} & 7.8$\pm$2.\rlap{2} & \mbox{~~}0.4$\pm$\mbox{~~}2.\rlap{0}  & 26.4$\pm$5.\rlap{5} & 79.5$\pm$2.\rlap{2} & 0.1$\pm$0.\rlap{2} & 0.0$\pm$0.\rlap{1} & 0.00$\pm$0.0\rlap{1} &  \mbox{~~}16.4$\pm$\mbox{~~}0.\rlap{8} & \mbox{~~}\mbox{~~}64.8$\pm$\mbox{~~}\mbox{~~}3.\rlap{2} & 0.35$\pm$0.03 \\
		Source 16 & 0.95$\pm$0.0\rlap{2} & 5.1$\pm$1.\rlap{6} & 27.4$\pm$\mbox{~~}7.\rlap{0}  & 20.4$\pm$8.\rlap{0} & 46.0$\pm$4.\rlap{9} & 0.7$\pm$0.\rlap{6} & 1.0$\pm$0.\rlap{4} & 0.50$\pm$0.0\rlap{7} &  \mbox{~~}\mbox{~~}3.9$\pm$\mbox{~~}0.\rlap{5} & \mbox{~~}\mbox{~~}\mbox{~~}5.7$\pm$\mbox{~~}\mbox{~~}0.\rlap{3} & 0.43$\pm$0.06 \\
		\hline
	\end{tabular}
	\label{CIGALEmodelsout}
\end{table*}

The fit of SAGE0536AGN shows that the emission from this object is not solely due to the AGN, as expected from the visible galaxy seen in survey images (see top left in Figure \ref{fig:Galfit}), $\sim$ 11\% is from the host galaxy. Extinction due to polar dust is the highest for SAGE0536AGN compared to the rest of the sample. Accretion power is smallest for SAGE0536AGN.

The fit of SAGE0534AGN however shows that the emission is almost solely due to the AGN. Compared to SAGE0536AGN, SAGE0534AGN shows similar $\tau$, $i$ and $R$ (radial thickness of torus) values. Extinction in polar dust is minimal compared to SAGE0536AGN. The opening angle is expected to be $\sim40^\circ$ from observations \citep[e.g.][]{2016MNRAS.458.2288S}. SAGE0534AGN has the expected opening angle, whilst SAGE0536AGN has the largest opening angle of the sample, implying a thinner torus.

\subsubsection{C\textsc{igale} models of the t-SNE sample}

The majority of sources show a lack of host galaxy contribution, $f\textsubscript{AGN} > 70$\%, which implies differences in the dusty torus (shape, density, etc.) are causing the differences. This is shown by the ranges of the other parameters such as $R$, $\tau$, $pl$, $q$ and polar dust extinction. All the sources, except Source 7, show an inclination angle between $0 < i < 45^\circ$, implying the central engine of the AGN is seen for all sources. All the sources, except Source 2, show $R>$ 20, implying a sample with a thick torus, some of which may be thicker than the models allow (10 $>R>$ 30). 

\subsection{X-ray observations}\label{xraysection}

Seven of these sources have been detected at X-ray 
energies with the XMM-Newton telescope \citep{2001A&A...365L...1J} (Table \ref{tab:Xrayfluxes}).

\begin{table}
\caption{The average, error-weighted 0.2--12 keV X-ray flux and the corresponding luminosity and their identifications from \protect\cite{2018Maitra}.}
    \centering
    \begin{tabular}{|l|c|c|l|}
        \hline\hline
        \llap{S}ource  & 0.2--12 keV flux  & Luminosity & Designation \\
        Name & (10$^{-13}$ erg s$^{-1}$ cm$^{-2}$) & (10$^{36}$ W) &  \\
       \hline
       6 & 0.76 $\pm$ 0.09 & 47.27$\pm$5.59 & \llap{J0}11408.02$-$723243.1  \\
       7 & 2.58 $\pm$ 0.14 & \mbox{~~}2.56$\pm$0.14 & \llap{J0}05551.53$-$733110.1  \\
       8 & 0.16 $\pm$ 0.07 & \mbox{~~}1.57$\pm$0.69 &  \\
       9 & 1.13 $\pm$ 0.11 & 58.49$\pm$5.69 & \llap{J0}12108.43$-$730713.1 \\
       10 & 0.50 $\pm$ 0.20 & \mbox{~~}0.59$\pm$0.24 & \\
       12 & 6.94 $\pm$ 0.30 & 64.85$\pm$2.80 & \llap{20}8.16034.100 \\
       13 & 1.72 $\pm$ 0.18 & 54.83$\pm$5.74 &  \llap{J0}05732.73$-$721302.1 \\
       \hline
    \end{tabular}
    
    \label{tab:Xrayfluxes}
\end{table}

Comparing with the C\textsc{igale} parameters calculated, the overall luminosity of the AGN increases with X-ray luminosity as expected. Also, the X-ray luminosities decrease with both the AGN inclination angle, polar extinction and AGN dust percentage, as expected, as when the central engine becomes more obscured, X-ray emission decreases. Overall, from looking at Table \ref{CIGALEmodelsout}, the X-ray luminosities of the sources with $f\textsubscript{AGN}<$99\% are in general lower than for $f\textsubscript{AGN}>$99\% sources. 


\subsection{The host galaxies}\label{galfitsection}

The host galaxy of SAGE0536AGN is resolved in VMC images, giving the appearance of a red galaxy. The fits of C\textsc{igale} also show that $\sim$ 11\% of this object's total dust emission is due to the host galaxy. C\textsc{igale} also calculated Sources 7, 8, 10, 15 and 16 to have emission contributed by the host galaxy. Of these, only Sources 10 and 16 have host galaxies that are resolved in the VMC images.

The appearance of the host galaxies provides insight into what step of evolution they are in, be they red dead elliptical, blue and star-forming, or intermediate as a green valley galaxy \citep{2014SerAJ.189....1S}.

\subsubsection{G\textsc{alfit}}

G\textsc{alfit} \citep{2002AJ....124..266P} is a well-known software used for galaxy decomposition and by using it we hoped to shed some light on the structure of the three host galaxies. It uses parametric functions to model objects as they appear in 2D images, i.e.\ modelling their light distributions. It can be used to determine the global morphology or to dissect a galaxy into its separate components such as bulge, disk, bar, etc.

\begin{figure}
	\includegraphics[trim={0.5cm 1.5cm 0cm 0cm},width=\columnwidth]{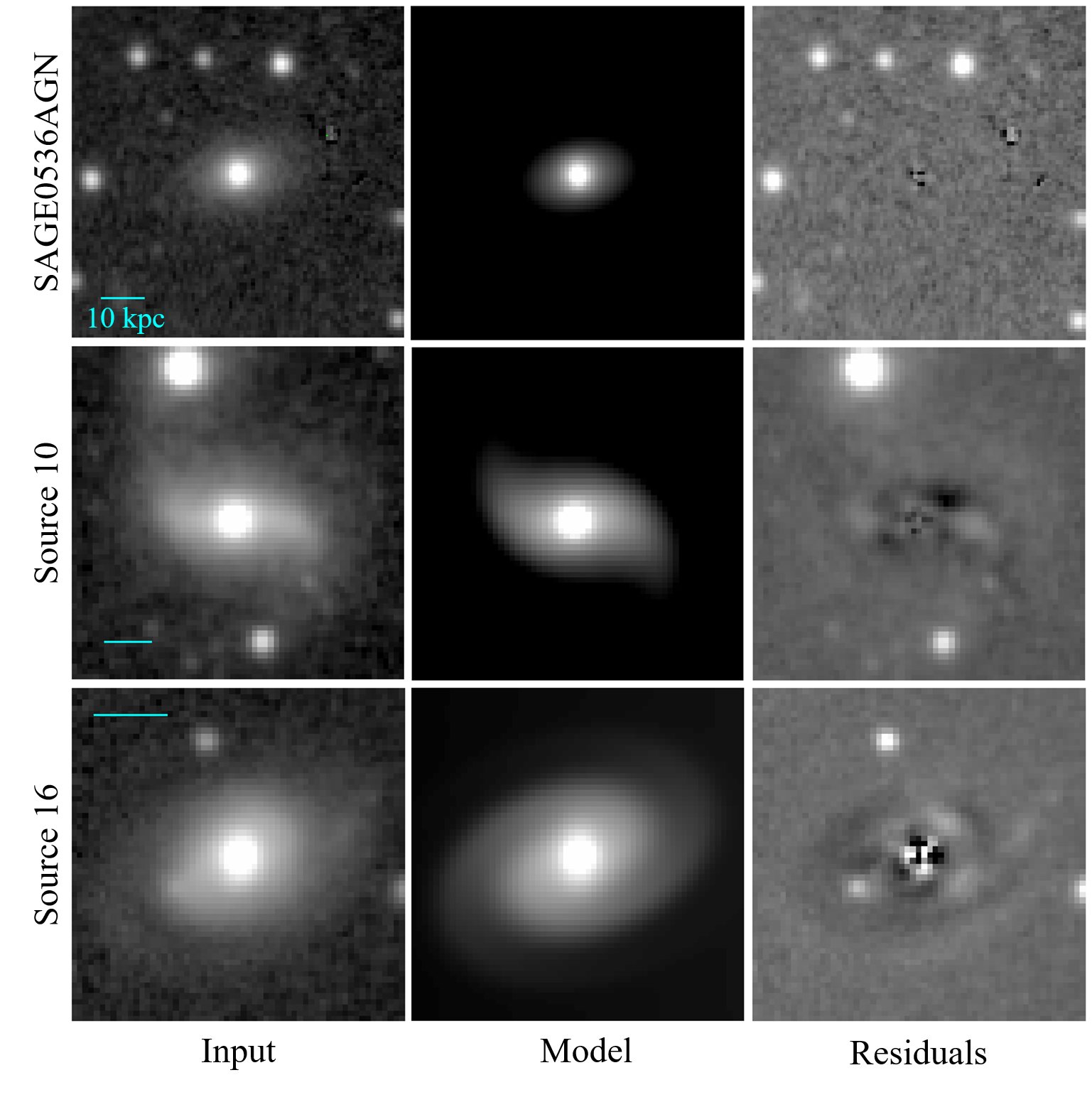}
    \caption{G\textsc{alfit} models of SAGE0536AGN (top), Source 10 (centre) and Source 16 (bottom). The cyan line represents 10 kpc based off of the respective redshift of each source. The input images are $K$\textsubscript{s} band VMC images.}
    \label{fig:Galfit}
\end{figure}

We used the S\'ersic profile function, as varying the S\'ersic exponent (which  determines the light profile) can match the other available functions in G\textsc{alfit}. The G\textsc{alfit} modelling was done using VMC $K$\textsubscript{s} band images. The models used are shown in Table \ref{Galfit_func}. Each AGN was fit with three S\'ersic functions, one for the host galaxy and two for the central component that includes the AGN. The models are shown in Figure \ref{fig:Galfit}.

\begin{table*}
    \centering
        \caption{The functions fitted to SAGE0536AGN, Source 10 and Source 16 in G\textsc{alfit} and their parameters. All sources were fitted with one S\'{e}rsic function for the host galaxy and two S\'{e}rsic functions for the central AGN. A sky background object was also fitted, the only parameter of that used was sky background at centre of fitting region [ADUs], fitted prior to fitting the other models by setting an estimate of the background and allowing G\textsc{alfit} to iterate and find the best value for the background. The magnitude is the total $K_{\rm s}$ Vega magnitude from the VMC survey. $R_{\rm e}$ is the effective radius in kpc (calculated from the redshift, conversion factors are 2.49 kpc arcsec$^{-1}$ for SAGE0536AGN, 3.34 kpc arcsec$^{-1}$ for Source 10 and 2.26 kpc arcsec$^{-1}$ for Source 16), such that half of the total flux is within $R_{\rm e}$. $R_{\rm e}$ for AGN is not meaningful since the AGN is not resolved. $n$ is the S\'{e}rsic exponent. $b/a$ is the axis ratio. The position angle is the angle the major axis, $a$, is orientated to. To account for the presence of spiral arms the S\'{e}rsic components of the host galaxy include PA rotation angle function. The bar radius is the radius where the rotation reaches roughly $20^\circ$. The 96\% asymptotic radius is the radius at 96\% tanh rotation. Rotation is the cumulative coordinate rotation out to the asymptotic radius. The asymptotic spiral arm powerlaw is related to the rotation, $\theta \propto r^{a}$, where $r$ is the radius and $a$ is the powerlaw.}
        \begin{tabular}{|l|l|l|l||l|l|l||l|l|l|l|l|} 
        \hline\hline
         &  \multicolumn{3}{c}{SAGE0536AGN} & & \multicolumn{3}{c}{Source 10} & & \multicolumn{3}{c}{Source 16} \\
         \hline
          Object type & AGN  & Bulge & Host & & AGN  & Bulge & Host & & AGN  & Bulge & Host \\
          Magnitude & 13.1 & 13.2 & 12.7 & & \mbox{~~}14.9 & 16.1 & \mbox{~~}14.5 & & \mbox{~~}14.6 & \mbox{~~}14.7 & \mbox{~~}\mbox{~}14.1\\
          $R_{\rm e}$ (kpc) & \mbox{~~}1.3 & \mbox{~~}2.3 & \mbox{~~}7.8 & & \mbox{~~}\mbox{~~}0.4 & \mbox{~~}1.1 & \mbox{~~}\mbox{~~}8.9 & & \mbox{~~}\mbox{~~}1.5 & \mbox{~~}\mbox{~~}3.4 & \mbox{~}\mbox{~~}11.6\\
          $n$ & \mbox{~~}0.6 & \mbox{~~}0.6 & \mbox{~~}0.6 & & \mbox{~~}\mbox{~~}0.2 &  \mbox{~~}1.7 & \mbox{~~}\mbox{~~}0.7 & & \mbox{~~}\mbox{~~}0.5 & \mbox{~~}\mbox{~~}0.7 & \mbox{~}\mbox{~~}\mbox{~~}0.7\\
          $b/a$ & \mbox{~~}0.9 & \mbox{~~}1.0 & \mbox{~~}0.7 & & \mbox{~~}\mbox{~~}0.6 & \mbox{~~}0.5 & \mbox{~~}\mbox{~~}0.4 & & \mbox{~~}\mbox{~~}1.0 & \mbox{~~}\mbox{~~}0.8 & \mbox{~}\mbox{~~}\mbox{~~}0.9\\
          Position angle (degrees) & $-$6.5 & 51.7 & 77.8 & & $-$55.3 & \mbox{~~}6.8 & \mbox{~~}86.9 & & $-$19.8 & $-$87.9 & \mbox{~}\mbox{~~}12.8\\
          PA rotation func. & none & none & none & & none & none & power & & none & none & power\\
          Bar radius (kpc) & - & - & - & & - & - & \mbox{~~}16.1 & & - & - & \mbox{~~}\mbox{~~}\mbox{~}3.9\\
          96 \% asymp. radius (pixels) & - & - & - & & - & - & 106.9 & & - & - & \mbox{~}\mbox{~~}96.0\\
          Rotation (degrees) & - & - & - & & - & - & 126.7 & & - & - & \mbox{~}100.4 \\
          Asymp. spiral arm powerlaw & - & - & - & & - & - & \mbox{~~}$-$3.3 & & - & - & \mbox{~~}$-$5.2\\
          Inclination to L.o.S. (degrees) & - & - & - & & - & - & \mbox{~~}\mbox{~~}0.0 & & - & - & $-$47.0\\
          Sky Position Angle (degrees) & - & - & - & & - & - & \mbox{~~}\mbox{~~}7.7 & & - & - & \mbox{~}\mbox{~~}67.5\\
        \hline
        \end{tabular}
    \label{Galfit_func}
\end{table*}

The host galaxies of Sources 10 and 16 display rotation as a function of radius, as is seen in spiral galaxies. G\textsc{alfit} allows for coordinate rotation in the light profile, and in this case we use the powerlaw spiral function in conjunction with the S\'{e}rsic function to account for the spiral arms. The residuals of SAGE0536AGN also suggest the presence of spiral arms, though they were not fit here.

When the S\'{e}rsic exponent, $n$, is large, it has a steep inner profile (cusp), and a highly extended outer wing. When $n$ is small, it has a shallow inner profile (core) and a steep truncation at large radius. For the host galaxies of these three AGN, SAGE0536AGN has $n=$ 0.62, implying a form between a Gaussian function ($n\sim$ 0.5) and an exponential disk ($n\sim$ 1). Source 10 has $n=$ 0.71 and Source 16 has $n=$ 0.70, implying host galaxies closer to exponential disk than SAGE0536AGN.

Combining the AGN and bulge components of each source provides a total magnitude brighter than the host galaxy, as expected. The AGN appears unresolved in the images, so the $R$\textsubscript{e} calculated by G\textsc{alfit} for the S\'{e}rsic profile are not true values. A S\'{e}rsic exponent, $n$ $\sim$ 0.5 gives a Gaussian component, and a Gaussian component with $R_{\rm e}\sim$ 0.5 pixels is an alternative for fitting a PSF profile and therefore an unresolved source, such as an AGN. Source 10 AGN has $R$\textsubscript{e} $\sim$ 0.5 pixels, showing it is as expected an unresolved source. SAGE0536AGN and Source 16 both have $R_{\rm e}\sim$ 1.5 pixels, implying the AGN bulge is slightly resolved.


All three galaxies are brighter at redder wavelengths, implying dust and/or lack of star formation. However, these three galaxies also have the appearance of spiral galaxies. This could imply a recent shut-down of star formation and that the galaxy has yet to transition morphologically into an elliptical galaxy. This could mean green-valley galaxies.

\section{Discussion}\label{discussionsection}

\subsection{IR properties and selection criteria}\label{selectionsection}

The infrared has proven to be an effective wavelength regime to select AGN in and therefore many selection criteria for AGN have been created for IR wavebands based on previously spectroscopically identified AGN.

\begin{figure}
	\includegraphics[trim={0.3cm 1.5cm 0.3cm 0.5cm},width=\columnwidth]{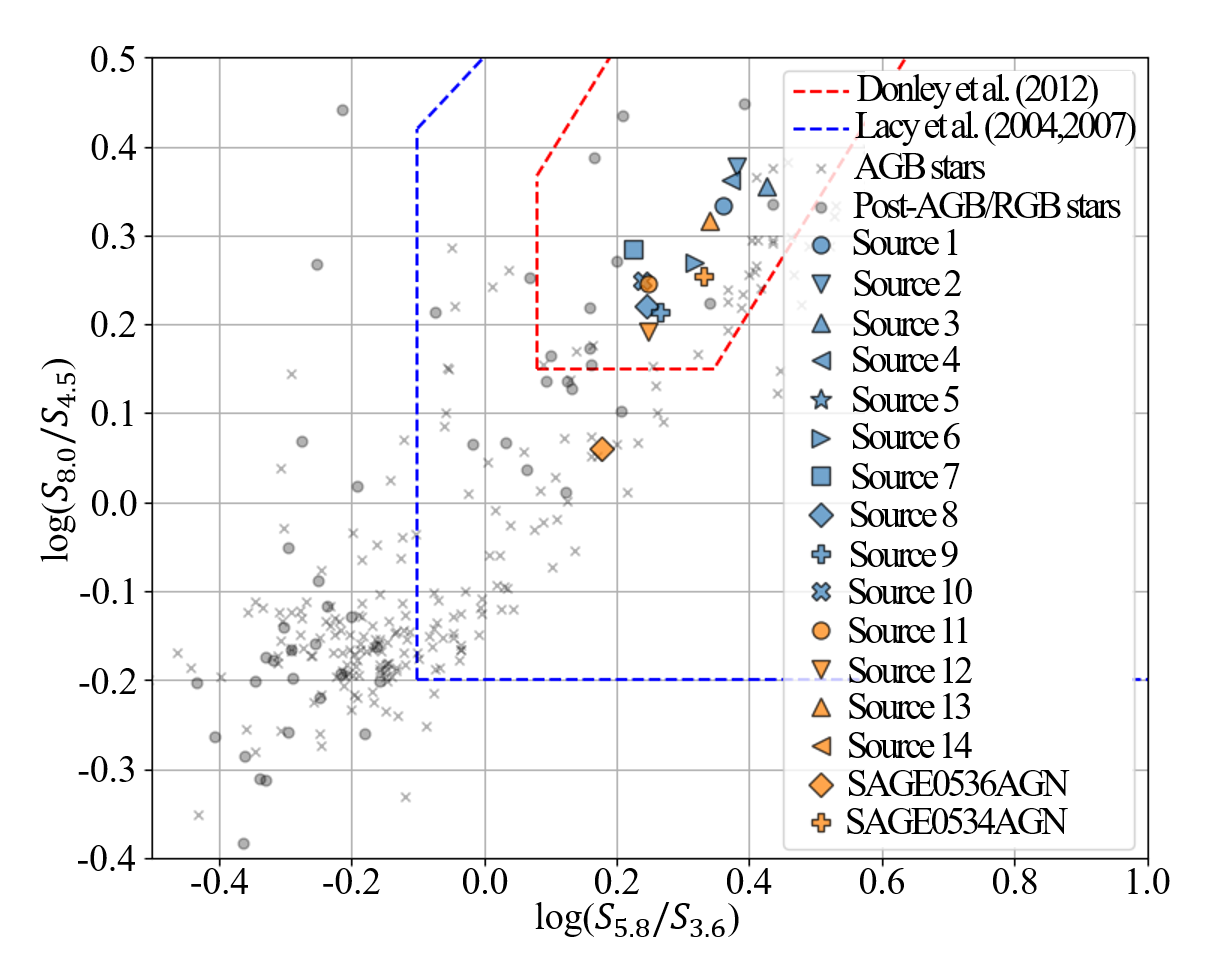}
    \caption{IRAC colour--colour diagram for the sample. The red and the blue dashed lines indicates the \protect\cite{2004Lacy} and \protect\cite{2012Donley} AGN selection criteria, respectively.}
    \label{fig:CC2}
\end{figure}

AGN selection criteria have been created for Spitzer IRAC and WISE wavebands by \cite{2004Lacy}, \cite{2005Stern}, \cite{2012Mateos} and \cite{2012Donley}. The \cite{2012Donley} wedge, shown in Figure \ref{fig:CC2}, was designed to be an improvement on the \cite{2004Lacy} and \cite{2005Stern} wedges, as it excludes high-redshift star-forming galaxies whilst incorporating the best aspects of the previous AGN selection wedges. All but SAGE0536AGN fall within the \cite{2012Donley} wedge. SAGE0536AGN, however, does still fall within the \cite{2004Lacy} wedge. This implies that sources similar to SAGE0536AGN could potentially be missed by the \cite{2012Donley} wedge.

\begin{figure*}
\centering
\begin{tabular}{c}
	\includegraphics[trim={0.8cm 1cm 0cm 0cm},width=\textwidth]{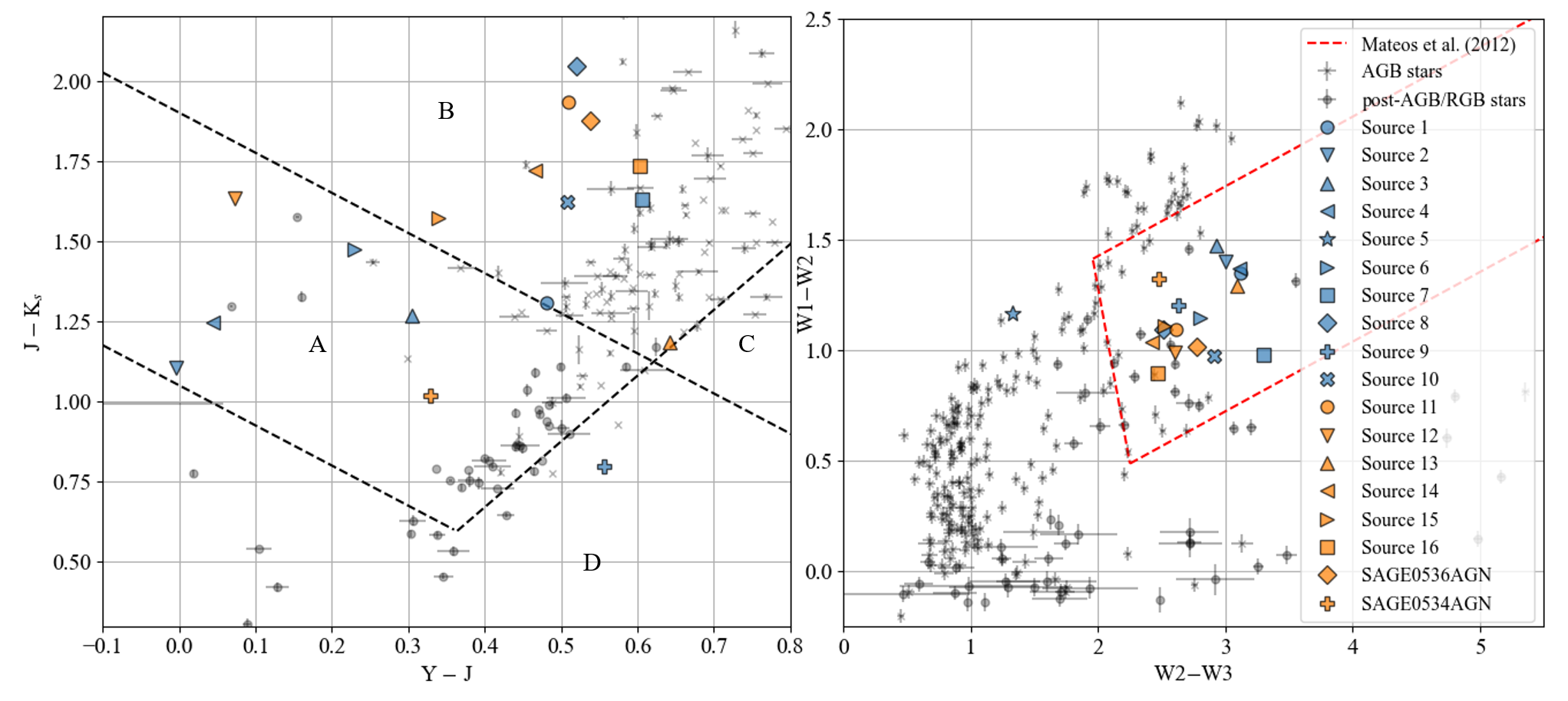}
	\end{tabular}
    \caption{ (left) VISTA colour--colour diagram of the sample. The regions A, B, C and D were devised by \protect\cite{2013Cioni}. Regions A and B are where most known AGN are found and are well matched by the models. Region C is dominated by reddened Magellanic Cloud sources and region D is populated by stars and low confidence AGN. (right) AllWISE colour--colour diagram of the sample. The sample is clumped together in the AGN region, apart from Source 5, a carbon star, which is to the left of AGN region and within the region populated by AGB stars. The red dashed line indicates the \protect\cite{2012Mateos} AGN selection criteria. Spectroscopically observed AGB stars in the Magellanic Clouds have been added to both plots as grey points and show how the two classes can be mistaken for the other.}
    \label{fig:VMC_CC2}
\end{figure*}

The \cite{2012Mateos} wedge, shown in the left panel of Figure \ref{fig:VMC_CC2} (right) shows that all sources fall well within the expected area in the colour--colour diagram. Also, as expected, Source 5 (the carbon star) falls outside this area. All the sources fall well within the criteria from \cite{2005Stern}, except SAGE0536AGN which is only just within its bounds.

\cite{2013Cioni} created AGN selection criteria which are shown in Figure \ref{fig:VMC_CC2} (left). It separates the colour--colour space into four regions. Regions A and B are where most known AGN are found where point-like AGN dominate region A and AGN with visible host galaxies dominate region B. The average redshift was found to be $z=$ 1.22$\pm$0.25 in region A and $z=$ 0.44$\pm$0.25 in region B. Region C was found to contain reddened Magellanic sources and region D was found to contain stars and low-confidence AGN. As expected, none of our sources are found in region C. Source 9, however, was found unexpectedly in region D, where stars dominate. The three sources with clear host galaxies, Source 10, 16 and SAGE0536AGN are found, as expected, in region B. All sources that have C\textsc{igale} fits with $f_{\rm AGN}<$ 0.99 are found in region B and all of the sources found in region A have $f_{\rm AGN}=$ 0.99. Sources 11, 13 and 14 are also found in region B, despite the predicted $f_{\rm AGN}=$ 0.99. Source 5 is found at $Y-J=$ 1.4 and $J-K_{\rm s}=$ 4.6 (not on the diagram), when it would be expected to be found in region C or D. From the sample the average redshifts for A and B are, $z\sim$ 1.02 and $z\sim$ 0.35, respectively, as expected.

AGB and post-AGB/RGB stars, classes known for being confused with AGN, have been added to the plots. These stars have all been spectroscopically observed \citep{1998A&A...332...25G,1998A&A...329..169V,1999A&A...346..805V,1999A&A...351..559V,2005A&A...438..273V,2006A&A...447..971V,2008A&A...487.1055V,2014MNRAS.439.2211K} and are all in the Magellanic Clouds. In Figure \ref{fig:CC2} the locus of the AGB stars is outside the two AGN selection criteria, however some are still found within the Lacy wedge, some of which are avoided with the Donley wedge. In Figure \ref{fig:VMC_CC2} (right) most of the stars are outside the Mateos wedge. Of those that encroach on the AGN criteria, AGB stars are in the top of the wedge, whilst post-AGB/RGB stars are at the bottom of the wedge. In Figure \ref{fig:VMC_CC2} (left) AGB stars can be mostly found in region B, whilst post-AGB/RGB stars can be found in region A. For all colour--colour diagrams AGB and post-AGB/RGB stars can be found amongst the AGN sample. It is known that combining near-IR and mid-IR selection techniques can efficiently select a high number of AGN \cite{2022arXiv220405219B}, and the combination of WISE and VMC colour selection techniques has the potential to efficiently remove the AGB and post-AGB/RGB sources.

\subsection{Green-valley}\label{greenvalleysection}

To determine whether the sources of the t-SNE selected sample are blue star-forming, green-valley or quiescent galaxies, they can be plotted on a diagram of star-formation rate (SFR) versus stellar mass of the host galaxy \citep[e.g. ][]{2016NatCo...713269C,2018MNRAS.477.3014B}.

\begin{figure*} 
\centering
\begin{tabular}{c}
	\includegraphics[trim={0.5cm 0.5cm 0cm 0cm},width=0.95\textwidth]{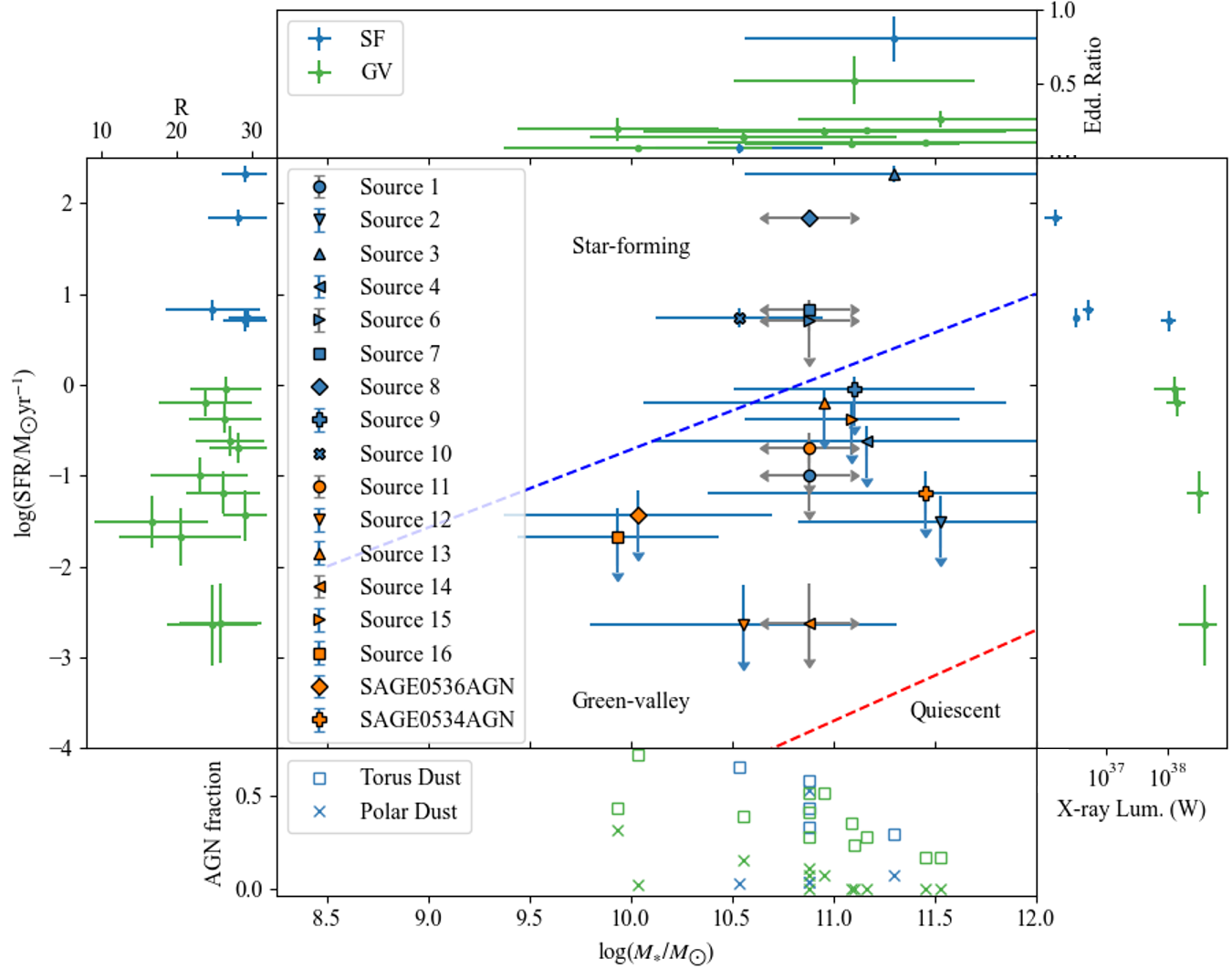}
	\end{tabular}
    \caption{(centre) The position of the t-SNE selected sample on a SFR versus stellar mass diagram. The blue and red dashed lines are an approximation of the boundary (at the 1$\sigma$ level in scatter from the main trend lines of the star-forming (SF) and quiescent galaxies) of the star-forming main sequence and the quiescent sequence, respectively, taken from \protect\cite{2016NatCo...713269C}. Those with blue error bars are sources with known black hole masses to calculate stellar mass from. Those with grey error bars are those with no known black hole mass, set at the average black hole mass of the sample. (top) How Eddington ratio changes with M\textsubscript{$\ast$}. (left) How $R$ changes with SFR. This shows that green-valley (GV) galaxies tend to have smaller tori than star-forming galaxies. (right) How X-ray luminosity changes with SFR. This shows that higher X-ray luminosities are seen in the green valley. (bottom) How the torus and polar dust emission fractions of the AGN change with M\textsubscript{$\ast$}. Accretion disk emission is the remaining fraction of AGN not plotted here.}
    \label{fig:greenval}
\end{figure*}

The SFR was calculated using the C\textsc{igale} best fits (after subtracting the AGN components) by using the correlation between total luminosity between 8 $\mu$m and 1000 $\mu$m and SFR as shown in \cite{Bell_2003}:
\begin{equation}
    SFR(M_{\odot}\rm yr^{-1}) = \begin{cases}
    1.57 \times 10^{-10}L_{\rm TIR}(1 + \sqrt{\frac{10^9}{L_{\rm TIR}}}) &, L_{\rm TIR} > 10^{11}\\
    1.17 \times 10^{-10}L_{\rm TIR}(1 + \sqrt{\frac{10^9}{L_{\rm TIR}}}) &, L_{\rm TIR} \leq 10^{11}\\
    \end{cases} 
\end{equation}
where $L$\textsubscript{\rm TIR} is the total luminosity between 8 $\mu$m and 1000 $\mu$m in solar luminosities. The stellar mass was calculated by using the correlation between black hole mass and stellar mass as described in \cite{2004ApJ...604L..89H}:
\begin{equation}
    \log(M_{\rm BH}) = -4.12 + 1.12\log(M_{\ast})
\end{equation}
where $M$\textsubscript{$\ast$} is the stellar mass. The resulting diagram can be seen in Figure \ref{fig:greenval} (centre). The C\textsc{IGALE} output for SFR was not used as most of the sources are dominated by the AGN, meaning the host galaxy, and therefore the SFR, could not be modelled accurately. For these sources the calculated SFR is an upper limit.

This plot shows that 12 out of 17 of the sources are green-valley galaxies. However, the far-IR fluxes from these galaxies are upper limits, meaning the SFR could possibly be lower and therefore in the quiescent region. SAGE0536AGN and Source 16 show spiral arm structure, meaning star formation shut off recently and they are at least more likely to be green-valley galaxies rather than quiescent.

Of the sources in the star-forming region, Source 6 may be a green-valley galaxy as its SFR is an upper limit. Of the other galaxies it is possible that the far-IR emission is not due to star formation. The far-IR from these galaxies could instead be accounted for by dust heated by the AGN beyond the torus. It has been shown that for torus opening angles of 20--70$^\circ$ \mbox{~} \citep{2018ApJ...862..118Z} the AGN emission will heat dust in the narrow-line region (where the polar dust is) if the black hole accretion disk is aligned with the galaxy plane \citep{2016ApJ...832....8B}, or the dust in the host galaxy if the accretion disk and the galaxy plane are misaligned \citep{2020A&A...638A.150V}.

We compare other observed characteristics of these sources and C\textsc{igale} model outputs with the distance along the evolutionary sequence, which we define as the distance of the source in the host mass versus SFR plane from the 1$\sigma$ scatter from the star-forming galaxies main sequence line.

From this we can see that the AGN fraction increases along the evolutionary sequence, which is expected as the host galaxy star formation reduces and the AGN becomes more prominent. Also as expected, the BH mass increases along the evolutionary sequence.

Source 3 is the furthest above the star-forming main sequence and has the largest Eddington ratio. The upper limit to Eddington ratio seems to decrease along the evolutionary sequence, this could imply its running out of fuel at later stages.

X-ray observed AGN are more likely to be found in the green valley than in the star-forming region \citep[e.g.][]{2009ApJ...693.1713T,2012A&A...541A.118P,2013ASPC..477..177P}. This is corroborated by the most X-ray luminous of the AGN being found in the green valley. X-ray luminosity increases along the evolutionary sequence, which could correspond with decreasing $R$ (ratio of outer torus radii to inner torus radii) and torus fraction, implying a thinning torus, which would mean that there is less dust and gas to absorb the X-ray emission.

\subsection{Radio analysis}\label{radiosection}

\subsubsection{Radio morphology}

\begin{figure*}
\centering
\begin{tabular}{c}
	\includegraphics[trim={0.5cm 1cm 0cm 0cm},width=\textwidth]{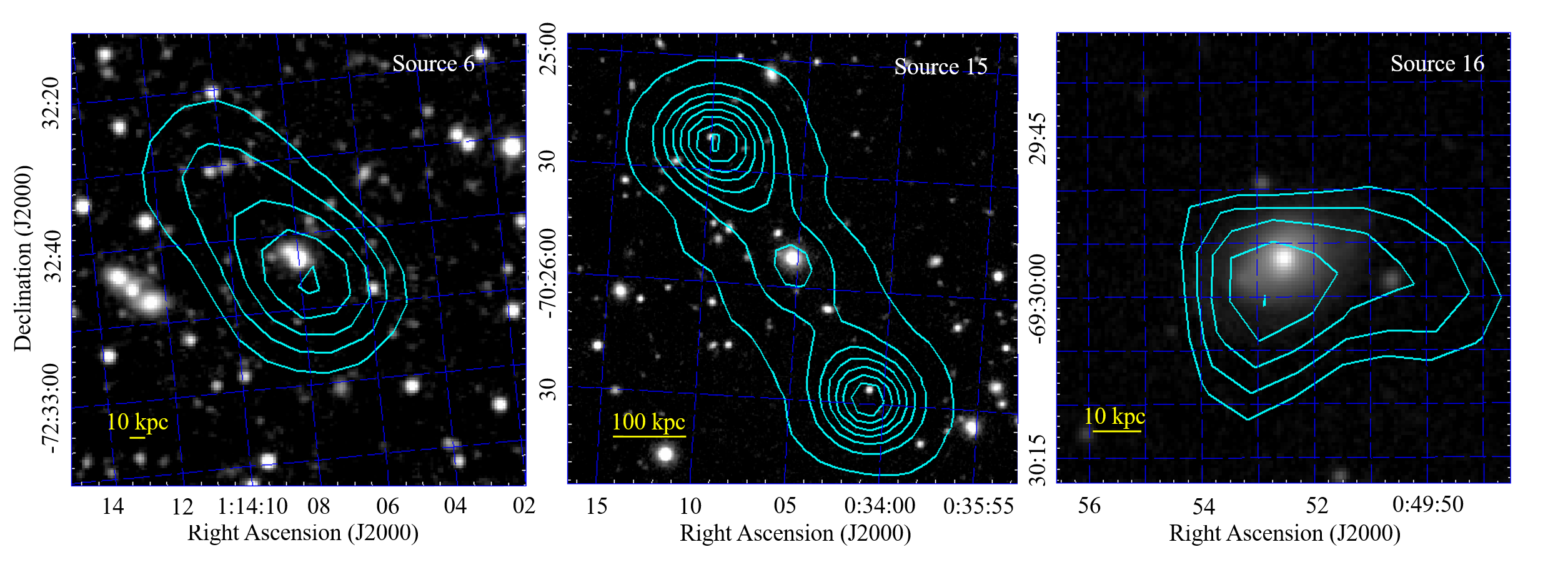}
	\end{tabular}
    \caption{ASKAP 1320 MHz radio flux of Sources 6, 15 and 16 shown as contours on top of the VMC K\textsubscript{s} band images. Source 6 has five linearly spaced contours from 0.18 -- 1.1 mJy. Source 15 has eight linearly spaced contours from 1 -- 70 mJy. Source 16 has five linearly spaced contours from 0.3 -- 0.72 mJy. Source 6 could potentially be an FR II source, with an offset radio peak and a counter-lobe at the other side. Source 15 is an FR II source. Source 16 radio emission extends towards the West most likely due to the point source to the West of it also being a radio source.}
    \label{fig:DAGN1320}
\end{figure*}

From the radio continuum images taken with ASKAP of the SMC and LMC, all 18 sources appear compact (unresolved at ASKAP resolutions) apart from three: Sources 6, 15, 16, which are shown in Figure \ref{fig:DAGN1320}. Sources 6 and 16's extended nature could be caused by radio emission from nearby sources blending with the main source.

Source 15's radio lobes would imply we are observing the source close to edge-on. The C\textsc{igale} model predicts $i\sim 0.4^\circ$, and therefore close to face-on. However, the radio image in Figure \ref{fig:DAGN1320} also shows a bright centre to the source. This could mean that the lobes are old relics, since then the source has rotated, and is now emitting a face-on radio jet.

\subsubsection{Spectral Indices}

We define the spectral index $\alpha$ by $F_\nu\propto\nu^\alpha$, where $F_\nu$ is the integrated flux density at frequency $\nu$. A flatter spectral index close to zero indicates free--free emission, and a steep negative spectral index of approximately $-0.7$, indicates synchrotron emission. Table \ref{tab:tspecInd} shows the radio luminosity and radio MHz spectral indices, $\alpha$. 

\begin{table}
\caption{Radio luminosities and spectral indices for the t-SNE selected sample from ASKAP (888 MHz for LMC, 960 MHz for SMC).}
    \centering
    \begin{tabular}{|l|c|c|c|}
        \hline\hline 
        Source Name & \multicolumn{2}{c}{L ($\times 10^{23}$ W)} & $\alpha$  \\
         & 888/960 MHz & 1320 MHz & \\
       \hline
       \llap{S}AGE0536A\rlap{GN} & \mbox{~~}\mbox{~~}0.5$\pm$\mbox{~~}0.1 & -- & -- \\
       \llap{S}AGE0534A\rlap{GN} & \mbox{~~}\mbox{~~}0.1$\pm$\mbox{~~}0.1 & -- & -- \\
       1 & \mbox{~~}\mbox{~~}0.1$\pm$\mbox{~~}0.1 & \mbox{~~}\mbox{~~}0.1$\pm$\mbox{~~}0.1 & $-$0.34  \\
       2 & \mbox{~~}99.7$\pm$14.2 & \mbox{~~}69.8$\pm$\mbox{~~}5.0 & $-$1.05 \\
       3 & 123.1$\pm$24.8 & \mbox{~~}77.8$\pm$\mbox{~~}7.6 & $-$1.46 \\
       4 & 179.8$\pm$18.0 & 142.1$\pm$\mbox{~~}7.2 & $-$0.69 \\
       6 & 115.1$\pm$\mbox{~~}9.6 & 109.9$\pm$\mbox{~~}5.3 &  $-$0.1 \\
       7 & \mbox{~~}\mbox{~~}0.6$\pm$\mbox{~~}0.2 & \mbox{~~}\mbox{~~}0.4$\pm$\mbox{~~}0.1 &  $-$1.12 \\
       8 & \mbox{~~}\mbox{~~}6.4$\pm$\mbox{~~}1.6 & \mbox{~~}\mbox{~~}3.6$\pm$\mbox{~~}0.6 & $-$1.73 \\
       9 & 113.9$\pm$10.4 & \mbox{~~}89.6$\pm$\mbox{~~}3.1 &  $-$0.69 \\
       10 & \mbox{~~}\mbox{~~}0.9$\pm$\mbox{~~}0.2 & \mbox{~~}\mbox{~~}0.9$\pm$\mbox{~~}0.1 & $-$0.38 \\
       11 & \mbox{~~}\mbox{~~}7.5$\pm$\mbox{~~}1.7 & \mbox{~~}\mbox{~~}5.9$\pm$\mbox{~~}0.5 & $-$0.67 \\
       12 & \mbox{~~}18.7$\pm$\mbox{~~}1.9 & \mbox{~~}11.4$\pm$\mbox{~~}0.8 & $-$1.58 \\
       13 & \mbox{~~}44.6$\pm$15.9 & \mbox{~~}35.1$\pm$\mbox{~~}6.4 & $-$0.93 \\
       14 & \mbox{~~}11.0$\pm$\mbox{~~}1.8 & \mbox{~~}\mbox{~~}6.5$\pm$\mbox{~~}0.6 &  $-$1.67 \\
       15 & -- & 340.3$\pm$\mbox{~~}4.8 & -- \\
       16 & -- & \mbox{~~}\mbox{~~}0.6$\pm$\mbox{~~}0.2 & -- \\
       \hline
    \end{tabular}
    
    \label{tab:tspecInd}
\end{table}

\begin{figure} 
	\includegraphics[trim={0.5cm 1.5cm 0cm 1cm},width=\columnwidth]{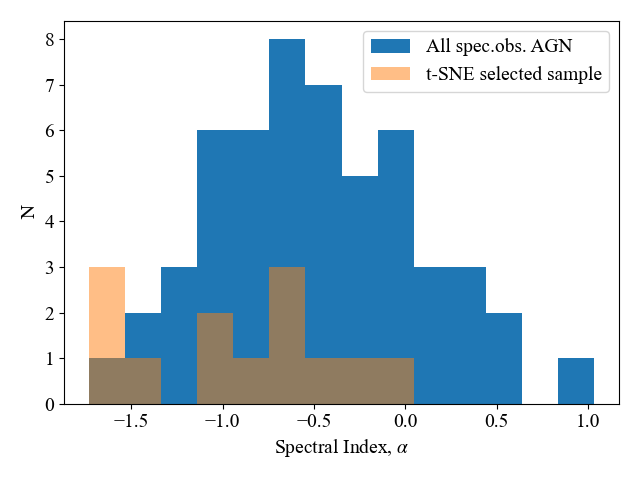}
    \caption{Spectral indices of the sample compared to other spectroscopically observed AGN behind the Magellanic Clouds.}
    \label{fig:SpecInd}
\end{figure}

Figure \ref{fig:SpecInd} shows the distribution of spectral indices for our sample compared to the spectroscopically observed AGN behind the Magellanic Clouds. This shows that whilst there is a peak at $\alpha \sim -0.7$, coinciding with the expected value for synchrotron radiation, there is also an unexpected peak at more negative values.

\subsubsection{Radio properties across the green valley}

\begin{figure} 
	\includegraphics[trim={0.3cm 1cm 0.4cm 1cm},width=\columnwidth]{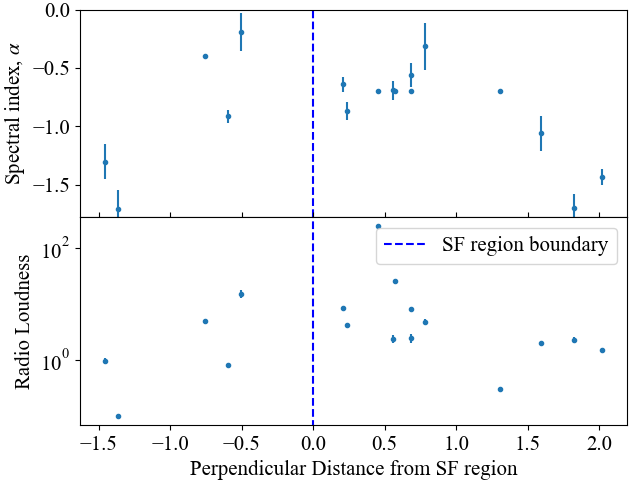}
    \caption{Variation of spectral indices, $\alpha$ (top), and radio loudness (bottom) of the sample as they transition from star-forming to green-valley galaxies.}
    \label{fig:radioev}
\end{figure}

We compare the radio properties of our sources with the distance along the evolutionary sequence, which we define as the distance of the source in the host mass versus SFR plane from the 1$\sigma$ scatter from the star formation main sequence line (Figure \ref{fig:radioev}).

From Figure \ref{fig:radioev} (top) we can see that the spectral index is steep at the start of the evolutionary sequence, is flattest at the beginning of the green-valley section and then steepens again. Figure \ref{fig:radioev} (bottom) shows that radio loudness, $R_{\rm AGN}$ (see Table \ref{CIGALEmodels}), is lowest at the start of the evolutionary sequence, reaches a peak at the beginning of the green-valley section, and then reduces again.

Sources 3 and 8 (two sources on the far left of Figure \ref{fig:radioev}) could be Compact Steep Spectrum (CSS) sources which are young sources that could go on to become large-scale Fanaroff--Riley II (FR II) objects \citep{1995A&A...302..317F}, such as Source 15. Furthermore, an observational signature of an AGN "switching-off" is also a steep spectrum ($\alpha < -1.5$). This is due to plasma ejected from the AGN losing energy causing high energy particles that radiate mostly at high radio frequencies to lose their energy fastest, making radio emission strongest at lower frequencies and causing a steep spectrum to be observed. This could imply that the AGN of the sources at the beginning of the evolutionary sequence have just switched on, explaining their steep radio spectrum. As the sources transition into the green valley the sources are at their radio loudest and have spectral indices $\sim-$0.7, implying steady synchrotron emission, after which the AGN, and subsequently the radio emission switches off, causing the radio loudness to decrease and the spectral index to steepen. The overall implication is that the AGN traces the transition from star-forming, across the green valley and into quiescence.


\subsection{AGN dust properties}\label{dustsection}

\subsubsection{Variability}\label{variabilitysection}

\begin{figure}
	\includegraphics[trim={0.5cm 1cm 0.5cm 1cm},width=\columnwidth]{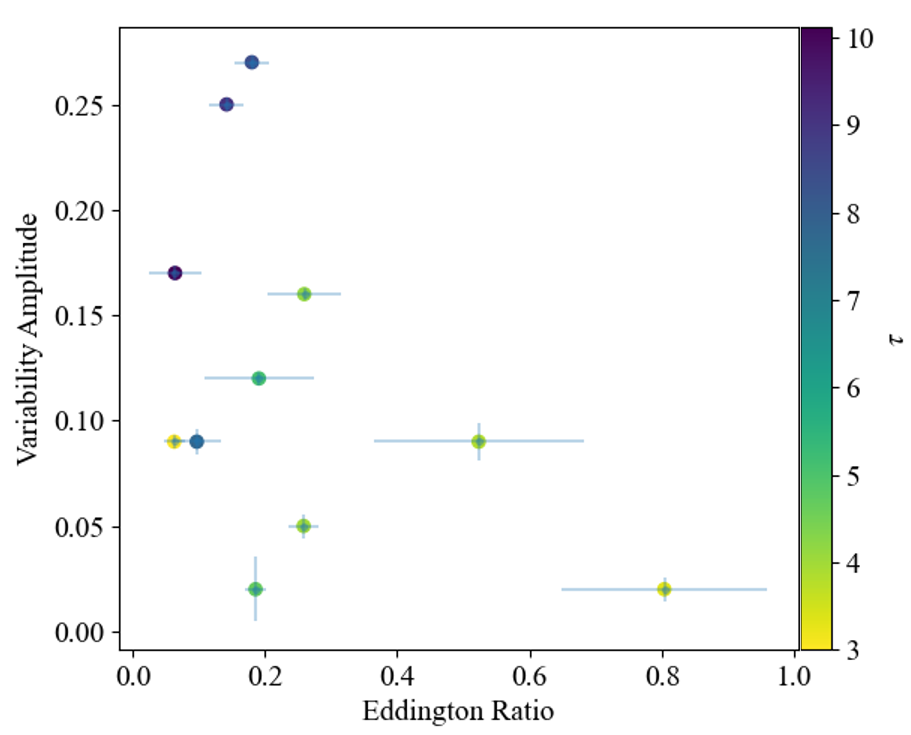}
    \caption{The comparison of variability with Eddington ratio and optical depth, $\tau$, of the t-SNE selected sample.}
    \label{fig:Eddvar}
\end{figure}

All the sources show some variability. From those with a known Eddington ratio we can see that high Eddington ratio sources tend to have little variability, while those with decreasing Eddington ratios tend to include sources with larger variability (Figure \ref{fig:Eddvar}). This relation holds true for black hole mass in place of Eddington ratio, so therefore smaller black holes (with smaller tori) vary the most whilst larger black holes (with larger tori) tend to vary less.

There is a general increase in variability with an increase in optical depth, implying the more emission from the accretion disk that is absorbed by the torus, the greater the variability. The Eddington ratio also decreases with increased optical depth and torus fraction. This implies Eddington ratio increases with less attenuation by the torus, as expected, but also that variability decreases with less attenuation by the torus, implying the torus is playing a part in the variability of the AGN that we observe. This could mean we are seeing variability in the attenuation of the emission from the accretion disk, instead of the variability of the accretion disk emission, which could mean high variability is caused by a "clumpy" torus moving around the accretion disk causing the amount of attenuation of the emission to increase and decrease.

High Eddington ratio sources vary the least. In general the highest Eddington ratios are at the start of the evolutionary sequence, whilst the lowest Eddington ratios are at the end of the green valley. This could imply that the sources at the beginning, where the AGN is just "switching on" and have the greatest amount of fuel and a high accretion rate, have the lowest variability, whilst the AGN that are starting to "switch off" and have the lowest amount of fuel and a lower accretion rate, have the highest variability. This could be due to the erosion of the dusty torus surrounding the AGN leading to a more porous torus and therefore increased variability. This would however be dependent on inclination angle of the AGN.

\subsubsection{C\textsc{igale} model components}\label{CIGALEmodelcompsection}

\begin{figure*}
\centering
\begin{tabular}{c}
	\includegraphics[trim={0.5cm 0.5cm 0cm 0cm},width=0.8\textwidth]{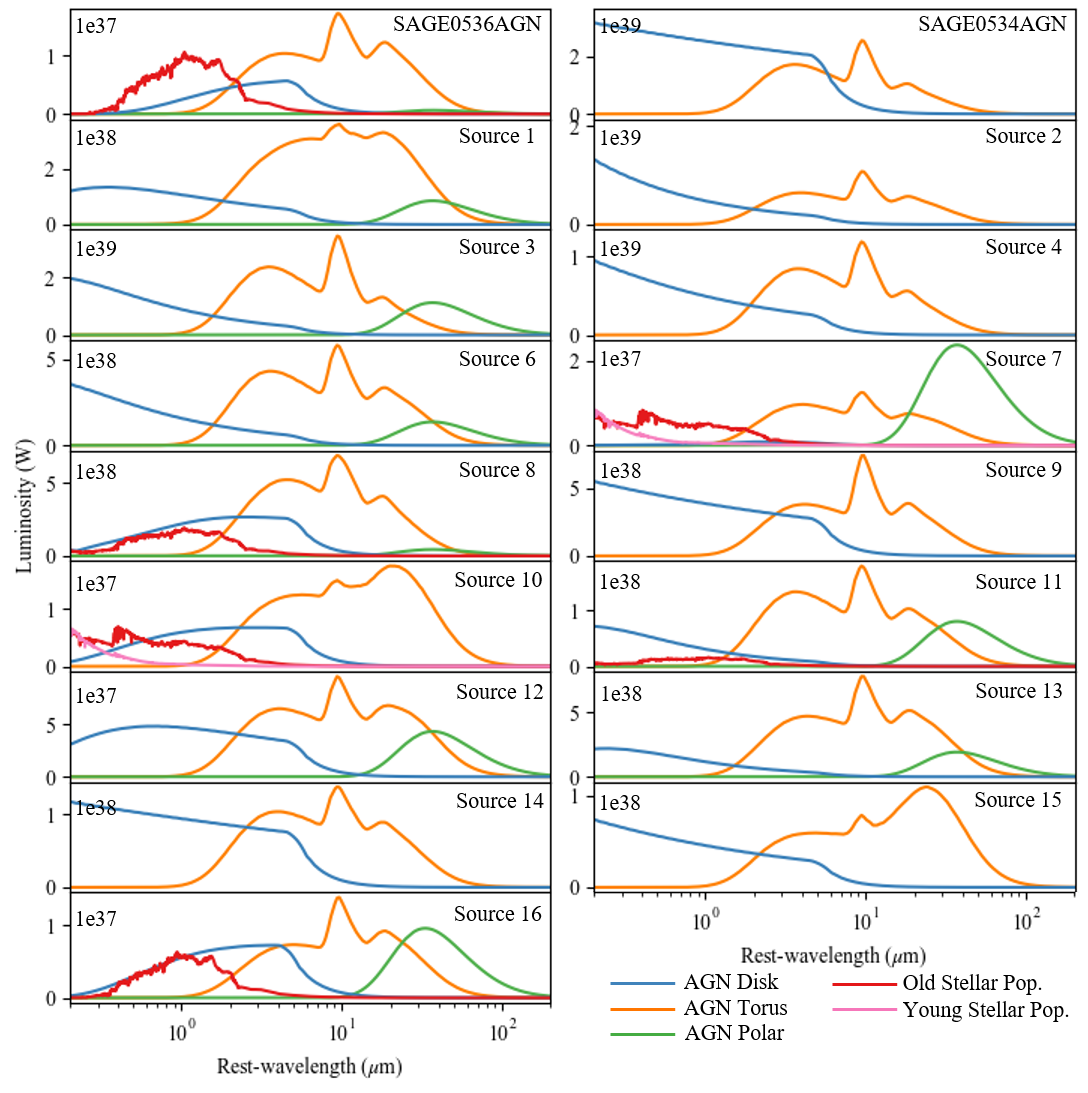}
	\end{tabular}
    \caption{Separate components of the C\textsc{igale} models for the t-SNE sample at rest-wavelength.}
    \label{fig:CIGALEplots}
\end{figure*}

C\textsc{igale} provides the separate models that make up the overall best-fit model. These models can be seen in Figure \ref{fig:CIGALEplots}. In the optical, the sources where the host galaxy dominates over the AGN disk are SAGE0536AGN and Source 7.

Polar dust contribution varies from source to source. Source 7 has the highest polar dust fraction at $\sim$ 53\%. This source also has one of the lowest AGN fractions of $\sim$ 72\%. However, the source with the lowest AGN fraction of $\sim$ 70\%, only has polar dust fraction of $\sim$ 3\%. Those with $<$ 1\% polar fraction can all be found in the green valley. These could be those where no outflows are present to push out the polar dust, and the AGN is starting to turn off.

\begin{figure}
	\includegraphics[trim={0.3cm 1cm 0.3cm 0.5cm},width=\columnwidth]{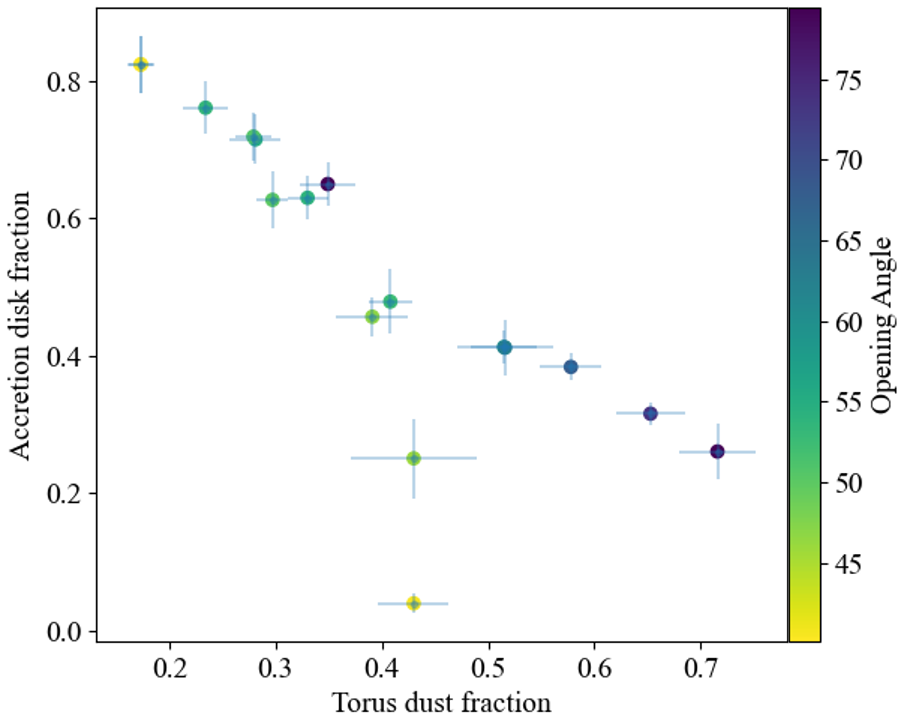}
    \caption{Accretion disk fraction versus torus disk fraction for the t-SNE selected sample. The colour bar represents the opening angle, $oa$. Sources 16 and 7 veer off from the negative correlation shown due to increased polar fraction, which also correlates with a decreased opening angle.}
    \label{fig:torusvsdisk}
\end{figure}

 As expected, we see a negative correlation between accretion disk fraction and torus disk fraction (Figure \ref{fig:torusvsdisk}), where SAGE0534AGN has the highest disk fraction and SAGE056AGN has the highest torus dust fraction, bracketing the sample. The opening angle also follows this trend, increasing with torus dust fraction (and decreasing with accretion disk fraction). However, Source 16, and to a larger extent, Source 7, veer off from these correlations due to their increased polar dust fraction. Their decrease in opening angle leads to an increase in polar dust fraction, which could be due to the increased space available at the poles with a smaller opening angle, as well as the poles being less obscured by the torus.

In general polar dust fraction increases with decreasing $R$ and $oa$, and increases with $i$. Note that Table \ref{CIGALEmodelsout} shows that Sources 7 and 16, that have the highest polar dust fraction, also have the highest values for $i$, which implies a link between the narrow-line region and polar dust fraction. The relation with $R$ could imply the presence of an outflow. This outflow would push out polar dust to become observable, increasing the polar dust fraction, as well as erode the dusty torus, decreasing $R$, which then in turn reveals more of the polar dust. The polar dust fraction would then also be expected to increase with $i$: as the accretion disk becomes more obscured by the dusty torus, then the torus dust and polar dust being pushed out by an outflow would become more prominent. As $oa$ decreases the view into the centre of the AGN opens, increasing the space over which polar dust can be found, thus increasing polar dust fraction.

\subsubsection{Silicate 9.7$\mu$m dust}\label{silicatesection}

The prominence and peak wavelength of the silicate emission of AGN varies. To quantify the strength of this emission we define it as the silicate peak relative to the continuum, at the wavelength where the silicate feature peaks \citep{2007ApJ...655L..77H}:
\begin{equation}\label{sileq}
    Si_{9.7\mu m}=\ln \frac{f_{9.7\mu m}(peak)}{f_{9.7\mu m}(continuum)}
\end{equation}

The silicate strength of SAGE0534AGN is calculated to be 0.24 $\pm$ 0.04 . In comparison, SAGE0536AGN yields a silicate strength of 0.85 $\pm$ 0.13.

C\textsc{igale} models the silicate dust as part of the AGN modelling. Calculating the silicate strength for all the sources as done previously with SAGE0536AGN and SAGE0534AGN revealed that silicate strength seems to increase with redshift. However, while the model correctly predicts that the silicate 9.7-$\mu$m feature is in emission for SAGE0536AGN and SAGE0534AGN, the model underestimates SAGE0536AGN (low-$z$ source) as Si\textsubscript{9.7$\mu$m} $\sim$ 0.54 and overestimates SAGE0534AGN (high-$z$ source) as Si\textsubscript{9.7$\mu$m} $\sim$ 0.72. This could imply the modelling of the silicate feature strength is not accurate or missing something.

The model predicts that all sources in this sample show silicate 9.7-$\mu$m emission. This could mean that the t-SNE selection separates those sources in emission from those in absorption. Confirming this requires follow-up mid-IR spectroscopy.

\subsubsection{Comparison to other silicate emitting AGN}\label{silicatecompsection}

SAGE0534AGN and SAGE0536AGN are not alone in their emission of silicate features. Comparing the two with more common less extreme versions of silicate emitting AGN may lend a clue to how these came about, whether they be extreme versions of an already established class of AGN or exist in a class of their own. To compare, a sample of local ($z <$ 0.1) type 1 AGN with silicate emission were taken from \cite{2020ApJ...890..152M}. They are a sample of 67 local ($z<$ 0.1) type 1 AGN. Another comparison was made with a sample from \cite{2014ApJ...788...98D} which include 46 2Jy radio galaxies  (0.05 $<z<$ 0.7) and 17 3CRR FRII radio galaxies ($z<$ 0.1) nuclei (AGN) with Spitzer spectra dominated by non-stellar processes. The sources in this sample have silicate strength calculated using Equation \ref{sileq}.

\begin{figure}
	\includegraphics[trim={0.3cm 1cm 0.3cm 0.5cm},width=\columnwidth]{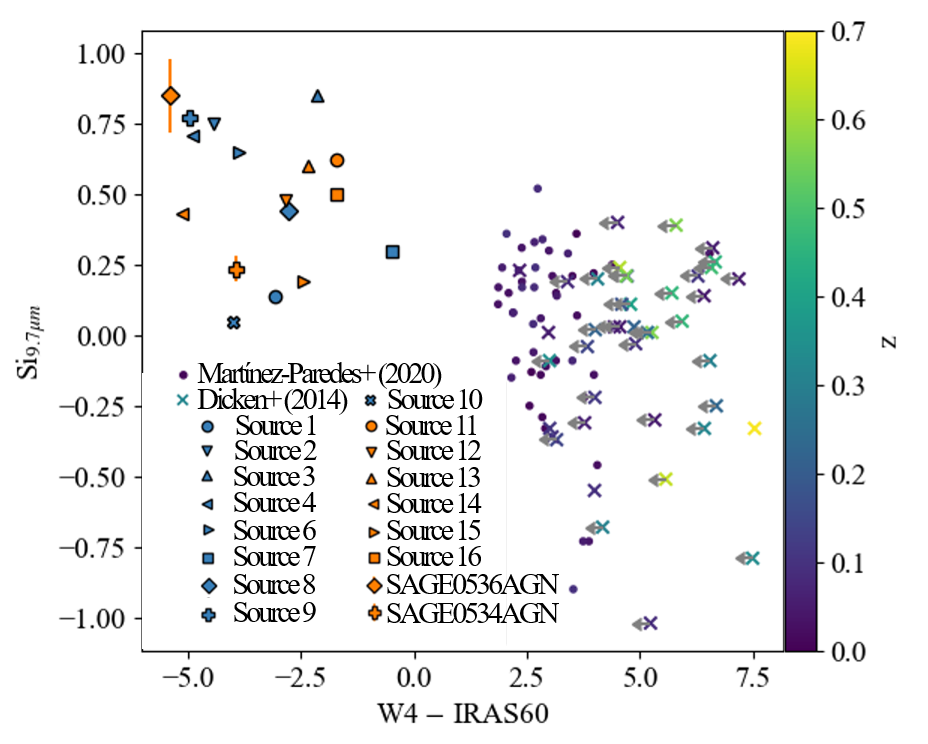}
    \caption{Silicate strength versus WISE band 4 -- IRAS 60-$\mu$m (in magnitudes). The IRAS60 measurements for SAGE0536AGN, SAGE0534AGN and the t-SNE sample are all calculated from the C\textsc{igale} model best fit.}
    \label{fig:si_strength}
\end{figure}

The silicate strength of SAGE0534AGN, SAGE0536AGN and the silicate emitting AGN sample were compared with their far-IR colour (WISE 4 (23 $\mu$m) -- IRAS 60 $\mu$m). For the sources in these samples that had no IRAS 60-$\mu$m measurements, a limit on the flux was calculated from the IRAS images. For the t-SNE selected sample the IRAS 60-$\mu$m magnitudes were estimated from the C\textsc{igale} best fit models. This comparison is shown in figure \ref{fig:si_strength}. The AGN sample are all to the right from SAGE0536AGN, SAGE0534AGN and the t-SNE selected sample. The higher redshift galaxies tend to be further to the right, but this could be because high-$z$ galaxies are most likely biased towards star-forming galaxies to make them bright enough. The limits on W4 -- IRAS60 could suggest there are already interesting sources observed. SAGE0536AGN however, remains apart due to its high silicate strength.

SAGE0536AGN is the strongest 10-$\mu$m silicate emitter currently known. In terms of torus properties predicted by C\textsc{igale}, SAGE0536AGN has one of the highest values for $R$, the largest values for $oa$ and torus fraction, lowest value for optical depth, $\tau$ and one of the lowest values for inclination angles. All this together could have provided the necessary environment for strong silicate emission. In contrast, SAGE0534AGN's silicate strength was overestimated. C\textsc{igale} predicts SAGE0534AGN has a similar inclination angle and optical depth to SAGE0536AGN, but the smallest values for $oa$ and torus fraction. Both of these sources also have very little polar dust, the presence of which correlates with weak or absent silicate emission \citep[e.g.][]{2020ApJ...892..149T}. This could imply that the increased silicate emission strength is due to a thicker torus with little to no polar dust to obscure the centre of the AGN. Of the rest of the t-SNE selected sample, the closest to SAGE0536AGN in terms of $oa$, torus fraction, polar fraction and inclination angle is Source 15, however the values for $\tau$ and $R$ are not as close. This could mean that the silicate emission of this source may rival that of SAGE0536AGN.

\section{Conclusions}

In this work we used unsupervised machine learning, t-SNE, with Gaia EDR3, VMC, AllWISE and EMU ASKAP photometric data, to find sources similar to SAGE0536AGN, the strongest 10-$\mu$m silicate emitter known, and SAGE0534AGN, a similar source with weaker silicate emission. This provided 16 sources to add to the sample. We took optical spectra of 15 of these sources and found that all but one were extragalactic in nature. From these spectra we calculated black hole masses and Eddington ratios. We used C\textsc{igale} to model the SEDs and characterise these sources, as well as used G\textsc{alfit} to model the morphology of the three nearest sources.

From this investigation we discovered most of the sources (12 out of 17) are in the green valley transitional phase, with the potential for some of these to be quiescent. We find that as sources move away from the star-forming phase and through the green valley phase the properties of the AGN change, such as the torus depletes, the Eddington ratio decreases, signalling the AGN is running out of fuel, and the X-ray luminosity increases as the material that would absorb it has depleted. Radio properties also change across this evolutionary sequence. The radio spectral slope starts off steep in the star-forming phase, before flattening to the expected value of $\alpha \sim -0.7$ at the beginning of the green valley, and then steepening again as the sources move further into the green valley. Radio loudness also follows this trend, starting off quiet in the star-forming phase, becoming loudest at the beginning of the green valley, before quieting again. This implies the "turning on" of the AGN to transition from star-forming to green valley, and then the AGN "turns off" again, before transition to quiescence.

All sources are variable and this variability decreases when there is less attenuation by the torus, implying the torus is playing a part in the variability.

SAGE0536AGN remains the most extreme 10-$\mu$m silicate emitter, which is not modelled well with C\textsc{igale}, which predicts weaker emission for SAGE0536AGN and stronger emission for SAGE0534AGN. C\textsc{igale} predicts all sources have silicate in emission. This needs to be verified by spectroscopic observations in the mid-IR, such as with the James Webb Space Telescope.

\section{Data Availability}

The data used to create the spectroscopy plots are available as online supplementary material. The inputs and outputs for CIGALE, Galfit and t-SNE are also available as online supplementary material. The spectroscopy data and the photometry data will be made available on the CDS\footnote{https://cds.u-strasbg.fr/} website when the paper is published.

The VMC photometry is available from ESO in the regular VMC data releases (\url{http://www.eso.org/rm/publicAccess#/dataReleases}). The VMC image data on the SMC, Bridge and Stream used in this paper are available in the VISTA Science Archive (VSA), at \url{http://horus.roe.ac.uk/vsa}. The deep stack VMC images of the LMC will be released mid-2022, whereas the individual observations are publicly available at the ESO archive (\url{http://archive.eso.org/cms.html}).

\section*{Acknowledgements}

We thank the anonymous referee for their feedback, which helped improve the paper. C.M.P.\ and J.E.M.C.\ acknowledge STFC studentship, J.O.A.\ acknowledges Nigerian Tertiary Education Trust funded studentship. A.N.\ acknowledges support from Narodowe Centrum Nauki (UMO-2020/38/E/ST9/00077). This paper uses observations made at the South African Astronomical Observatory (SAAO) under programme Pennock-2019-05-74-inch-257, and with the Southern African Large Telescope (SALT) under programmes 2021-1-SCI-018 (PI: Jacco van Loon), 2021-1-SCI-029 (PI: Jacco van Loon), 2021-1-SCI-032 (PI: Jacco van Loon) and 2021-2-SCI-017 (PI: Joy Anih). This paper also made use of spectra contributed by Makoto Kishimoto, observed with the European Southern Observatory's 3.6m telescope with EFOSC2 under programme 073.B-0501(A). This research made use of Astropy,\footnote{\url{http://www.astropy.org}} a community-developed core Python package for Astronomy \citep{2013A&A...558A..33A, 2018AJ....156..123A}. We have made extensive use of the SIMBAD Database at CDS (Centre de Données astronomiques) Strasbourg, the NASA/IPAC Extragalactic Database (NED) which is operated by the Jet Propulsion Laboratory, CalTech, under contract with NASA, and of the VizieR catalog access tool, CDS, Strasbourg, France. This project has received funding from the European Research Council (ERC) under the European Union’s Horizon 2020 research and innovation programme (grant agreement no. 682115). We thank the Cambridge Astronomy Survey Unit (CASU) and the Wide Field Astronomy Unit (WFAU) in Edinburgh for providing the necessary data products under the support of the Science and Technology Facility Council (STFC) in the UK.





\bibliographystyle{mnras}
\bibliography{ref.bib} 




\appendix



\bsp	
\label{lastpage}
\end{document}